\documentclass[a4paper,fleqn,usenatbib]{mnras}

\usepackage{txfonts}
\usepackage[T1]{fontenc}
\usepackage{ae,aecompl}

\usepackage{graphicx}	% Including figure files
\usepackage{amssymb}	% Extra maths symbols
\usepackage{times}
\usepackage{booktabs}

\input epsf.sty
\newif\ifAMStwofonts
\AMStwofontstrue

\title[Multi-phase environment of Centaurus~A]{The multi-phase environment in the centre of Centaurus~A}

\author[Borkar et al.]
       {A.~Borkar,$\!^{1}$\thanks{E-mail: abhijeet.borkar@asu.cas.cz} T.~P. Adhikari,$\!^2$\thanks{E-mail: tek@iucaa.in} A.~R\'o\.za\'nska,$\!^3$
       A.~G.~Markowitz,$\!^{3,4}$ P.~Boorman,$\!^{1,5}$
       \newauthor
       B.~Czerny,$\!^{6}$ G.~Migliori,$\!^{7,8}$ B.~De~Marco,$\!^3$ and V. Karas$^1$\vspace{0.3cm}\\
       $^1$ Astronomical Institute, Czech Academy of Sciences, Bo\v{c}n\'{\i} II 1401, 141\,00 Prague, Czech Republic\\
       $^2$ Inter University Centre for Astronomy and Astrophysics (IUCAA), Post Bag 4, Pune University Campus Ganeshkhind,\\
       Pune -- 411007, Maharashtra, India \\
       $^3$ Nicolaus Copernicus Astronomical Center, Polish Academy of Sciences, Bartycka 18, 00-716 Warsaw, Poland \\
       $^4$ University of California, San Diego, Center for Astrophysics and Space Sciences, 9500 Gilman Dr, La Jolla, CA 92093-0424, USA \\
       $^5$ Department of Physics \& Astronomy, University of Southampton, SO17 1BJ, Southampton, United Kingdom \\
       $^6$ Center for Theoretical Physics, Polish Academy of Sciences, Al. Lotnikow 32/46, 02-668 Warsaw, Poland \\
       $^7$ Dipartamento di Fisica e Astronomia, Universit\`{a} di Bologna, viale Berti Pichat 6/2, 40127 Bologna, Italy\\
       $^8$ INAF - Istituto di Radioastronomia, Via P. Gobetti, 101 40129 Bologna, Italy}

\date{Accepted XXX. Received YYY; in original form ZZZ}

\pubyear{2020}
\begin{document}
\label{firstpage}
\pagerange{\pageref{firstpage}--\pageref{lastpage}}
\maketitle

\begin{abstract}
We study the multi-phase medium in the vicinity of the active galactic nucleus Centaurus~A (Cen~A). Combined high-resolution observations with the ALMA and {\it Chandra\/} observatories indicate that the hot X-ray emitting plasma coexists with the warm and cold media in Cen~A. This complex environment is a source of CO lines with great impact for its diagnostics. We present the images from the two above-mentioned instruments covering the nuclear region (diameter of $10\arcsec$ i.e., $\sim$180 pc), and we study the conditions for plasma thermal equilibrium and possible coexistence of cool clouds embedded within the hot X-ray emitting gas. Further, we demonstrate that the multi-phase medium originates naturally by the thermal instability (TI) arising due to the interaction of the high-energy radiation field from the nucleus with the ambient gas and dust. We demonstrate that cold gas clouds can coexist in the mutual contact with hot plasma, but even colder dusty molecular clouds have to be distanced by several hundred pc from the hot region. Finally, we propose a 3-D model of the appearance of the hot plasma and the CO line-emitting regions consistent with the {\it Chandra\/} image and we derive the integrated emissivity in specific molecular lines observed by ALMA from this model. To reproduce the observed images and the CO line luminosity the dusty shell has to be $\sim$420 pc thick and located at $\sim$1000 pc from the centre.
\end{abstract}


%%%%%%%%%%%%%%%%%%%%%%%%%%%%%%%%%%%%%%%%%%%%%%%%%%%%%%%%%%%%%%%%%%%%%%%
\section{Introduction}

Centaurus A is a nearby active galactic nucleus (AGN) hosted in the galaxy NGC~5128 (hereafter we refer to the AGN and the innermost regions of NGC~5128 collectively as Cen A). Its redshift $z$ is 0.00183, and it is located at a distance $D=3.8\pm 0.1$~Mpc~\citep{harris2010}. Being the nearest radio-loud AGN, Cen~A has played a major role in our understanding of the nature of accreting supermassive black holes (SMBH), circumnuclear environments, connections between AGN and host galaxies, and jet ejection and emission processes. Our knowledge of Cen~A thus forms a template for studying AGNs at farther distances, including radio-loud AGN and other gas-rich ellipticals. Cen~A is also considered to be a prototypical FR~I object \citep{fanaroff74}.

Key insights have come from high-spatial resolution imaging with e.g., \textit{Chandra X-ray Observatory}, very long baseline interferometry (VLBI), and the Atacama Large Millimeter/Submillimeter Array (ALMA) and include e.g., small-scale spectral and kinematic behaviour of jet components \citep[e.g.][]{tingay2001,hardcastle2007,worrall2008,goodger2010,Sni19}; interactions of jet ejecta and the interstellar medium \citep{kraft2003,croston2009}; hot interstellar medium and non-hydrostatic motions \citep{kraft2008}; interaction of jet ejecta with stars or gas clouds in the host galaxy \citep{hardcastle2003,tingay2009}; identification of numerous individual X-ray point sources, including X-ray binaries \citep{burke2013}; and morphology and kinematics of the circumnuclear molecular gas, including disk-like structures, at the scales of tens to hundreds of parsecs \citep{neumayer2007,espada2009,espada2010}. For a review of the host galaxy properties, nucleus, radio lobes, and links to suspected past major mergers, see e.g. \citet[][]{israel1998}.

A key part of forming this knowledge-template is discerning the ionisation structure, dust content, and gas-dust interactions, particularly given the co-spatiality of the X-ray-emitting plasma and cold molecular clouds. The presence of an AGN provides positive and negative feedback, creating a complex interstellar environment. NGC~5128 possesses a dust lane which provides a large gas reservoir from which gas can be funnelled towards the central black hole to fuel accretion. The central engine in turn influences the central region with outflows and the jet which drives the gas away, which can lead to formation of a multi-phase environment in the vicinity of the SMBH.\@

In this paper, we study the condition for the thermal instability (hereafter TI) and see how it shapes the multi-phase gas in a region 10$\arcsec$ ($\sim$180~pc) in diameter centered on the core. Given its distance $D$ from us, the nucleus of Cen~A can be observed with high spatial resolution, thus giving us an insight into the plasma conditions under the combined influence of an active nucleus and a compact stellar cluster. We have analysed data observed with \textit{Chandra} and ALMA, resulting in images that enable us to study \textendash{} at high spatial resolution \textendash{} the hot diffuse plasma overlapping cold molecular gas in the innermost 10$\arcsec$ diameter.

To model the condition of TI, we use photoionization calculations with the {\sc cloudy} code\footnote{www.nublado.org}, version 17.01 \citep{Ferland2017}, which calculates the transfer of radiation through the matter for a wide range of temperature and density, and for different spectral shapes of illuminating radiation. We follow the approach presented by us in the cases of the Sagittarius~A* (Sgr~A*) SMBH at the centre of the Milky Way \citep{czerny2013c, kunneriath2014}, and in the ultra-compact dwarf galaxy M60-UCD1 \citep{rozanska14, rozanska17}.

The structure of this paper is as follows. In Sec.~\ref{sec:obs} we describe the current knowledge of the main properties of Cen~A based on the large collection of data from different instruments. In the same section, we present the ALMA and \textit{Chandra} high-resolution images used in this paper to follow the distribution of matter. Furthermore, we discuss the radiation field and luminosities collected from available observations of the nucleus of Cen~A, needed as input for our photoionization modeling. The set-up of the model including the description of parameters used in our numerical calculations is presented in Sec.~\ref{sec:model}. All results are presented in Sec.~\ref{sec:res}. The comparison to observations and potential ways to generalize our scenario are discussed
and concluded in Sec.~\ref{sec:dis}.

\begin{table*}
  \centering
  \begin{minipage}{170mm}
    \caption{Details of ALMA observations taken during 2015--2016 as part of the Cycle 3 observational campaign under project 2015.1.00483.S. See Sec.~\ref{subsec:alma} for further details.}\label{ALMAOBS}
    \begin{tabular}{lllll}
      % \hline\hline
      \toprule
      & \multicolumn{2}{c}{Band 3} & \multicolumn{2}{c}{Band 6} \\
      % \hline
      \midrule
      Date & 2015 Dec 26, 2016 Jan 13 & 2016 May 26 & 2016 Jan 10, August 16 & 2016 Jan 15, 16 \\
      Number of antennas & 34 (main array), 11 (ACA) & 34 & 44 & 46\\
      Baseline length (m) & 15--310, 8--43 & 15--640 & 15--1500 & 15--331 \\
      Primary beam ($\arcsec$ (pc)) & 53, 91 (954, 1600) & 55 (990) & 23 (414) & 23 (414)\\
      Synthesized beam ($\arcsec$) & 2.62 $\times$ 2.06, 15.96 $\times$ 8.49 & 1.4 $\times$ 1.3 & 0.68 $\times$ 0.62 & 1.25 $\times$ 0.97 \\
      Line rest frequency & 115.27 (CO v = 0, J = $1 - 0$)  & 110.20 (13CO v = 0 J = $1 - 0$) & 265.88 (HCN v = 0 J = $3 - 2$) & 271.98 \\
      (GHz)& & 109.78 (C18O v = 0 J = $1 - 0$) & 267.56 (HCO+ v = 0 J = $3 - 2$) & (HNC v = 0 J = $3 - 2$)\\
      & & 97.98 (CS v = 0 J = $2 - 1$) & & \\
      Spectral resolution & 122 kHz (320 m s$^{-1}$) & 122 kHz (332 m s$^{-1}$) & 244 kHz (275 m s$^{-1}$) & 244 kHz (269 m s$^{-1}$) \\
      Calibrators (Amplitude, & Ganymede/Mars, J1427$-$4206 & Ganymede & Ganymede & Callisto \\
      bandpass, phase) & J1321$-$4342/J1352$-$4412 & J1427$-$4206, J1321$-$4342 & J1427$-$4206, J1352$-$4412 & J1427$-$4206, J1352$-$4412 \\
      % \hline
      \bottomrule

  \end{tabular}
  \end{minipage}
\end{table*}

\begin{table}
  \begin{center}
    \caption{\small CO ($1 - 0$) observed luminosity from four regions, listed in the first column and shown in Fig.~\ref{fig:regions}. The second column lists the array used;  the third column lists the measured integral flux density $S_{\rm CO}$; the last column lists the line luminosity $L_{\rm CO}$ obtained using the standard conversion \citep{solomon97}.}
\label{tab:almareg}
  \begin{tabular}{lrll}
  \toprule
  Region & Array & $S_{\rm CO}$ [Jy km s$^{-1}$] & $L_{\rm CO}$ [erg s$^{-1}$]  \\
  \midrule
  Reg 1 & 12-m main & $3.0 \times 10^4$ & $1.97 \times 10^{38}$ \\
  Reg 2 & 12-m main & $5.8 \times 10^4$&  $3.85 \times 10^{38}$ \\
  Reg 3 & 12-m main & $6.9 \times 10^4$&  $4.59 \times 10^{38}$ \\
  Reg 4 & 7-m ACA & $7.3 \times 10^5$  &  $4.86 \times 10^{39}$ \\
  \bottomrule
  \end{tabular}
  \end{center}
\end{table}

\begin{figure*}
\centering
\includegraphics[scale=1.25]{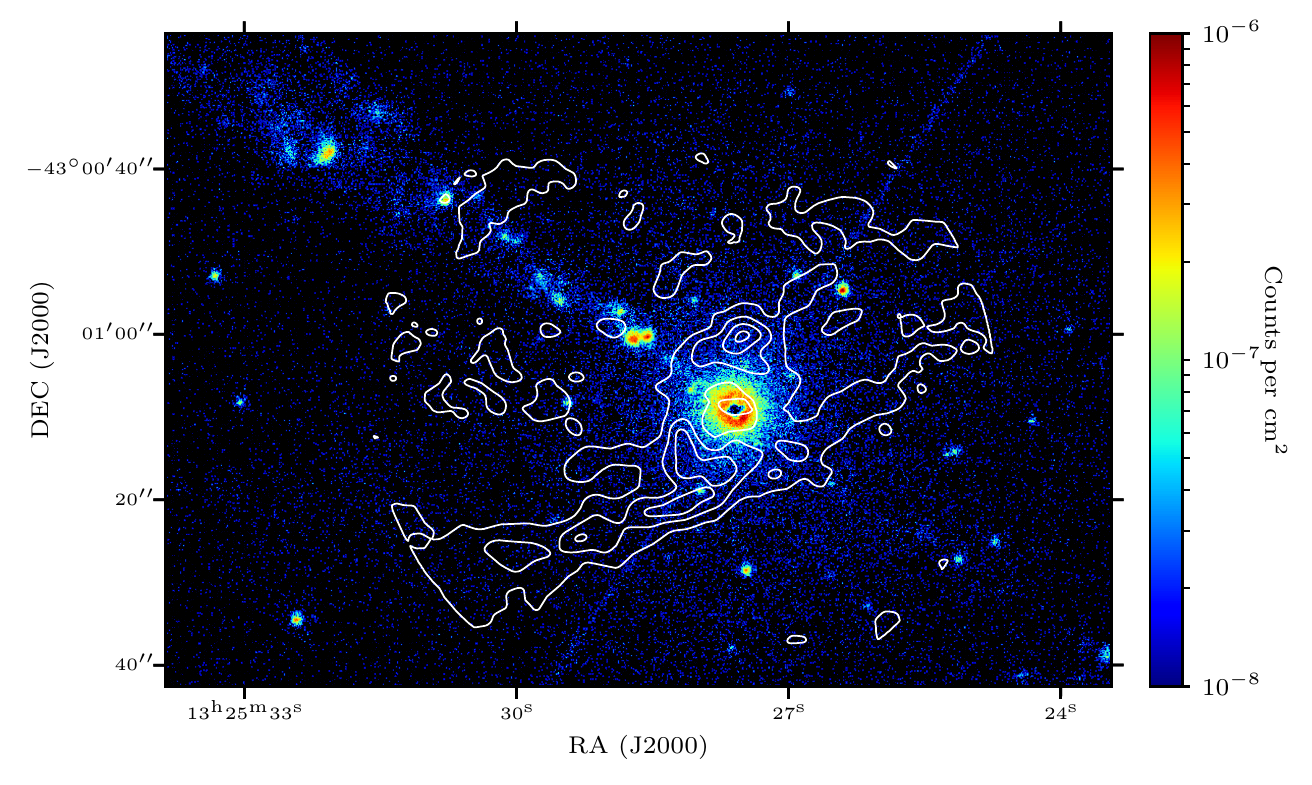}
\includegraphics[scale=1.3]{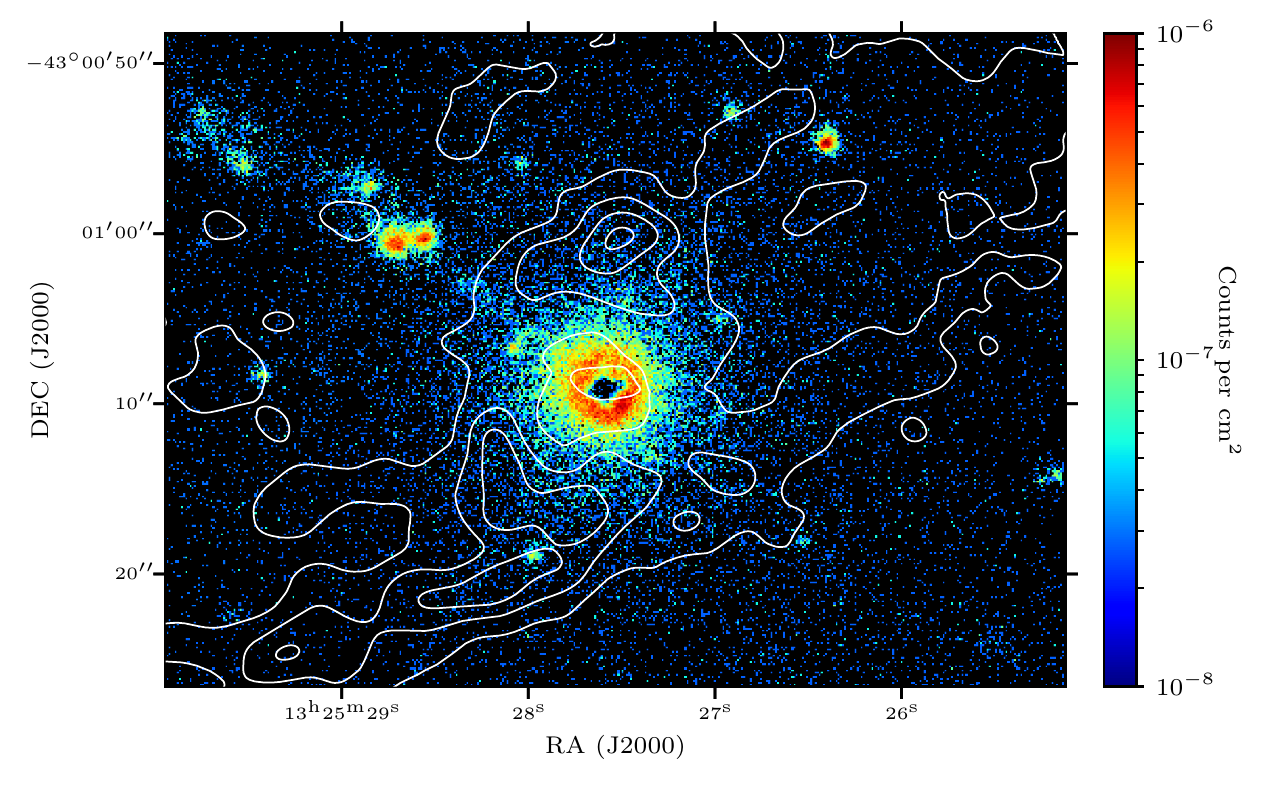}
\caption{{\it Chandra\/} X-ray images together with ALMA contours. The top image shows the central 1\,kpc of Cen~A in X-rays, with overlaid integrated CO ($1 - 0$) flux density contours from the ALMA main array. The contour levels are (4, 7.66, 11.33, 15)\,Jy beam$^{-1}$ km s$^{-1}$. The bottom image shows the same, but focused on the central 200 pc region of interest for this work.}
\label{fig:contour}
\end{figure*}

\begin{figure}
\includegraphics[scale=0.99]{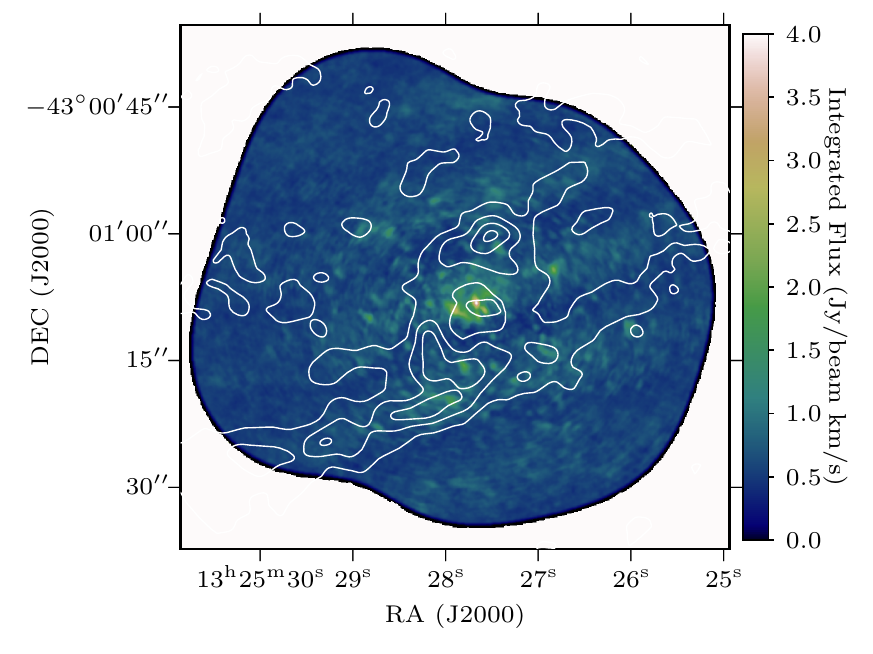}
\caption{Cen~A in the HCO+ ($3 - 2$) \textendash~267.557633 GHz line. Contours indicate CO ($1 - 0$) emission (same as in Fig.~\ref{fig:contour}).}
\label{fig:Hlines}
\end{figure}

\begin{figure}
\includegraphics[scale=0.99]{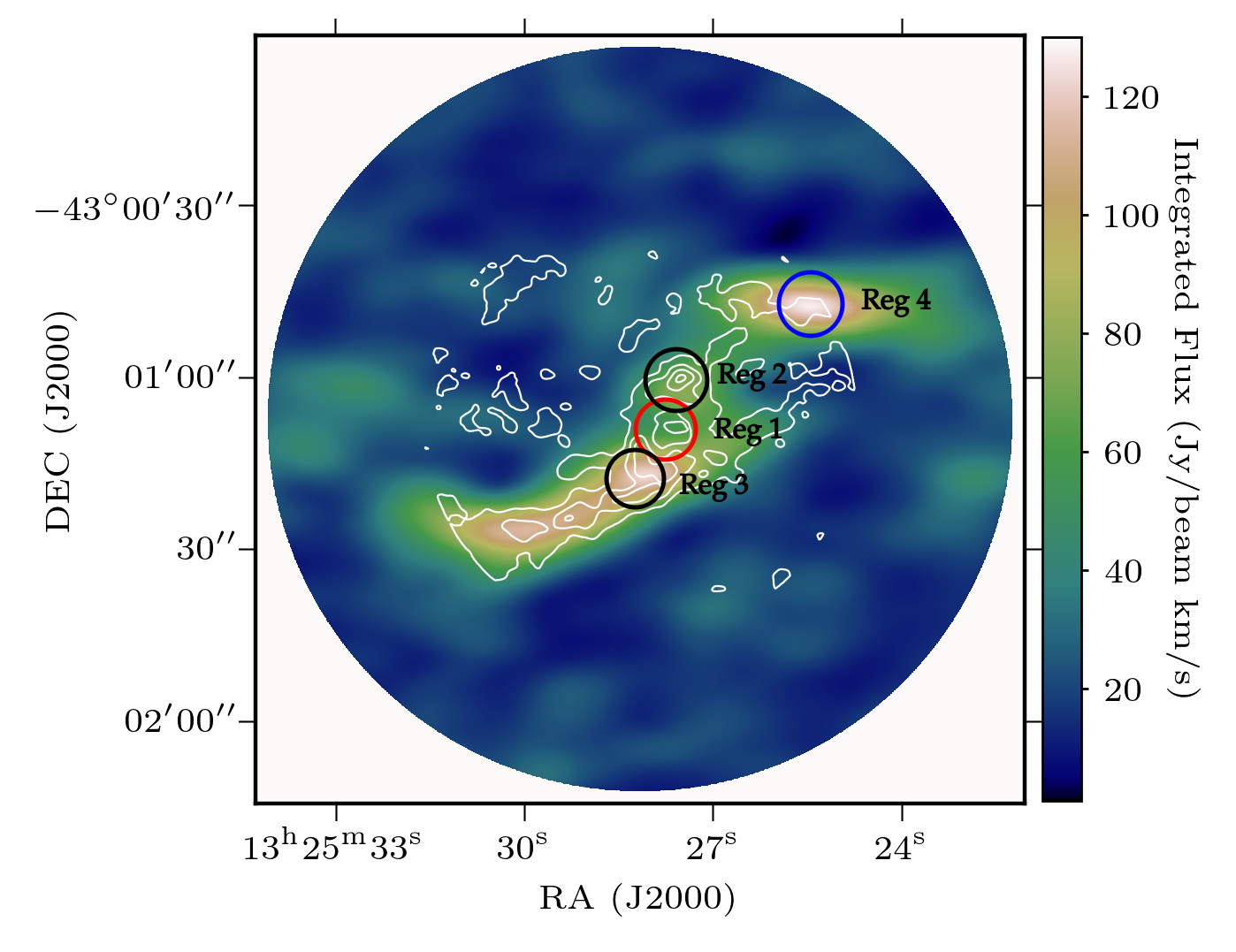}
\caption{Regions marked in the map for which the CO luminosity was calculated. Reg 1 is the central region, while Reg 2 and Reg 3 are regions of the same size taken by the 12-m ALMA main array. The most luminous region of the same size is detected by the 7-m ACA, and marked as Reg 4.}
\label{fig:regions}
\end{figure}

\begin{figure}
  \includegraphics[scale=0.5]{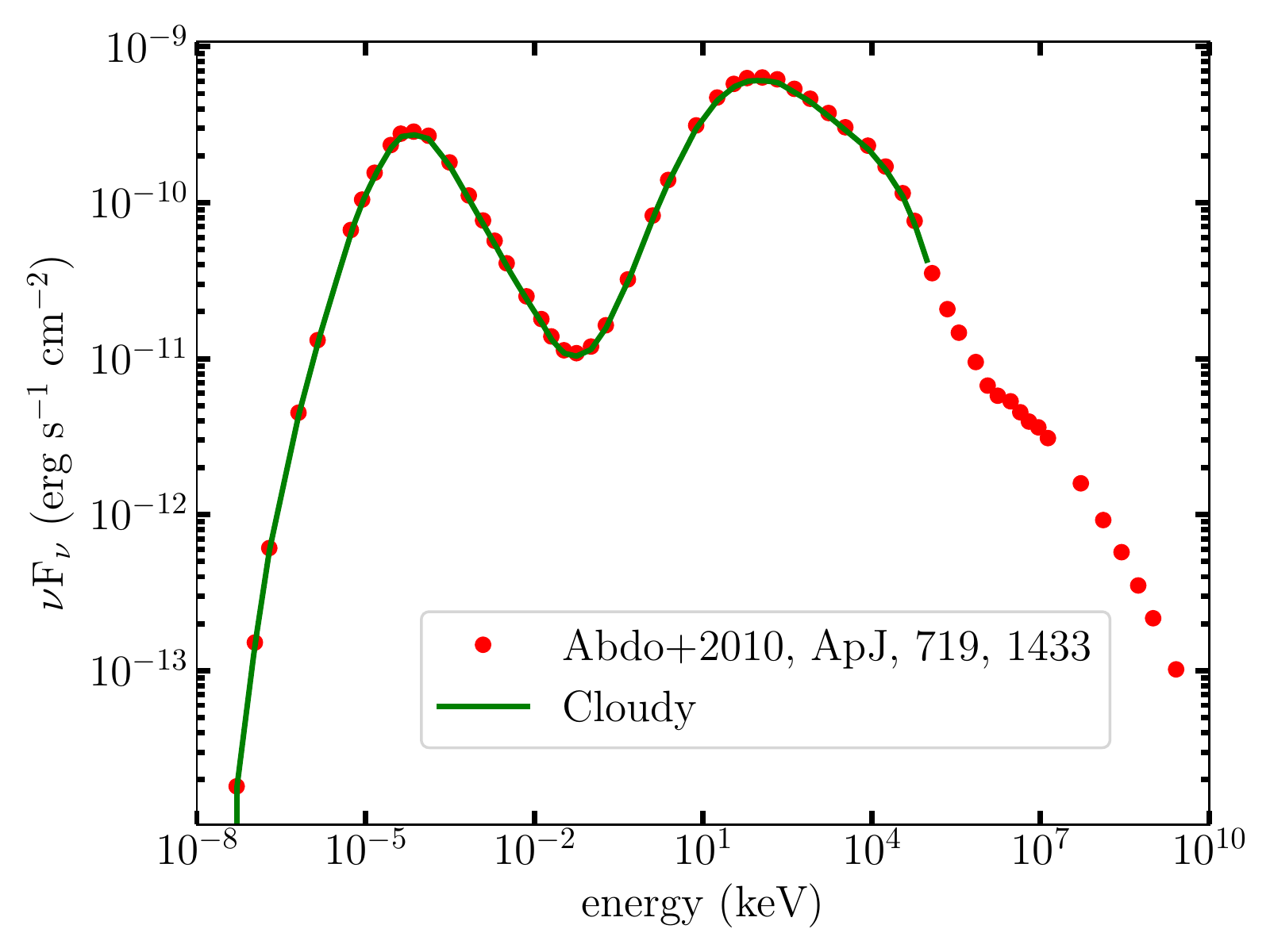}
  \caption{The spectral energy distribution (SED) of the nucleus of Cen~A from multi-wavelength observations, given by points. The solid curve denotes the SED used in this paper for photoionization calculations with the {\sc cloudy} code.}\label{cenA:sed}
\end{figure}

\begin{figure*}
\includegraphics[scale=0.90]{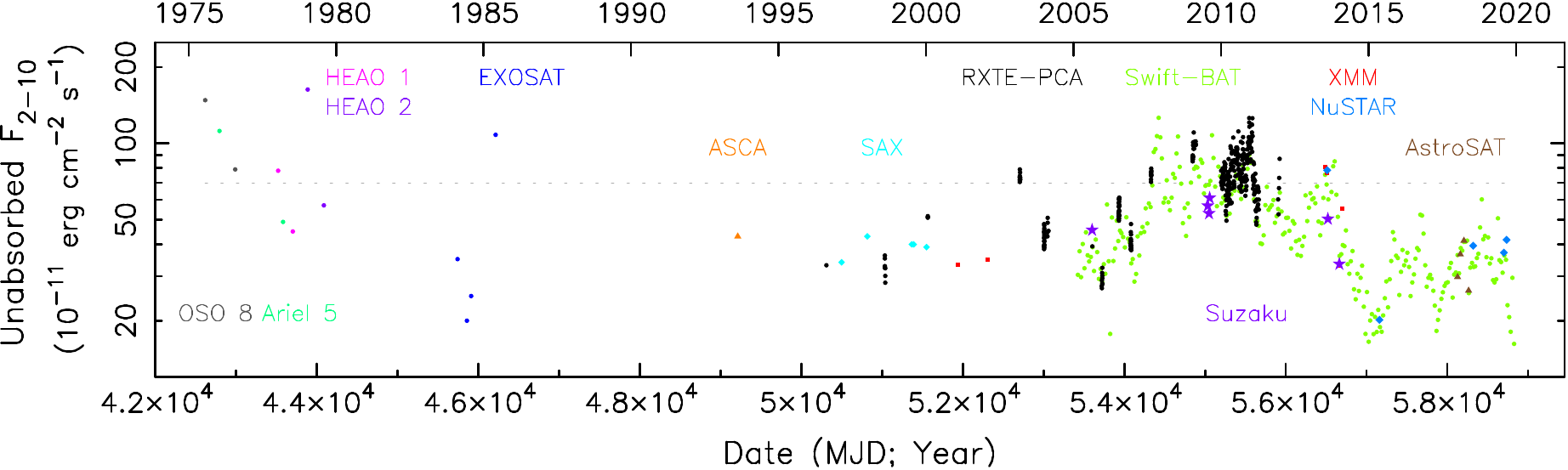}
  \caption{Historical unabsorbed $2 - 10$ keV power-law flux; references are listed in Section~\ref{data:nustar}. The dashed line is the long-term mean, $7.9\times10^{-10}$ erg cm$^{-2}$ s$^{-1}$, which, for a luminosity distance of 3.8 Mpc, corresponds to a 2--10 keV luminosity of $1.4\times10^{42}$ erg s$^{-1}$.}
\label{fig:lightcurve}
\end{figure*}

\section{Observations}
\label{sec:obs}

\subsection{Main properties of Cen~A}
\label{sec:mainp}

The mass of the central black hole of Cen~A determined by using neutral hydrogen gas kinematics spans typically best-fit values of $4.5-11  \times 10^7$~M$_{\odot}$ \citep{marconi2006, krajnovic2007, neumayer2010} and by using stellar kinematics spans values of $\sim 5.5 - 20 \times 10^7$~M$_{\odot}$ \citep{silge2005, cappellari2009}.

The observed activity of the Cen~A nucleus is rather low. The Bondi accretion rate was estimated to be $\dot{M} = 6.4 \times 10^{-4} M_{\odot}$~yr$^{-1}$ \citep{evanskraft2004}. A number of long ($\sim$100 ks) \textit{Chandra} exposures have allowed us to distinguish several features such as an extended X-ray jet \citep{hardcastle2007, worrall2008, goodger2010, Sni19}, the radio-lobe shock \citep{croston2009}, and numerous individual X-ray sources including X-ray binaries \citep{burke2013}.
The extended diffuse X-ray-emitting gas which extends up to 6 kpc also pervades the nuclear region \citep{kraft2008}.

ALMA observations have revealed a complex molecular gas structure in this gas-rich elliptical galaxy, consisting of a circumnuclear disk, and molecular arms stretching out to the dust lanes \citep{mccoy2017}. The study of emission lines has allowed the determination of the kinematics of the cold medium, suggesting some infall towards the nucleus \citep{espada17}.

To study the physical processes in the central region of Cen~A, we need both high-resolution images in different energy bands displaying spatial distribution of hot and cold gas around nucleus, and the broadband spectral energy distribution (SED) of the emission from the innermost region. The high-resolution images provide the estimation of distances at which hot and cold emission is detectable and where the emitting material is located. A realistic source SED is used as an input for the photoionization calculations as described in Sec.~\ref{sec:model}. Furthermore, to properly estimate the total cooling and heating due to all radiative processes, including photoionization, an accurate estimation of the value of bolometric luminosity is required, which is taken into account by the {\sc cloudy} code. All those properties of Cen~A are collected in this section.

\subsection{ALMA millimetre data}
\label{subsec:alma}

To follow the distribution of the nuclear matter in the vicinity of the SMBH, we used ALMA band 3 and 6 observations of central region of Cen~A, obtained as part of a Cycle 3 observation campaign under project 2015.1.00483.S (PI:\@ F.\ Israel). The observations were carried out between December 2015 and August 2016 with multiple executions of the schedule blocks with both the ALMA 12-m main array and the 7-m ALMA compact array (ACA). The main array observations were made with 34--46 antennas, while the ACA used 11 antennas, providing unprojected baseline lengths spanning 15--1500 metres for the main array and 8--43 metres for the ACA.\@ The primary beam (field of view) is obtained for the 12-m antennas is $53\arcsec$ ($\sim$950 pc) for band 3 and $23\arcsec$ ($\sim$410 pc) for band 6, and $91\arcsec$ ($\sim$1.6 kpc) for the 7-m antennas (band 3 only).

The observations were aimed at detecting line emission from $^{12}$CO, $^{13}$CO, C$^{18}$O, CS, HCN, HCO+ and HNC molecules in the central region of Cen~A. The CO ($1 - 0$) ($\nu_{\textup{rest}} = 115.27$ GHz), $^{13}$CO ($1 - 0$) ($\nu_{\textup{rest}} = 110.20$ GHz), C$^{18}$O ($1 - 0$) ($\nu_{\textup{rest}} = 109.78$ GHz), and CS ($2 - 1$)
 ($\nu_{\textup{rest}} = 97.98$ GHz) line emission was detected in band 3, and HCN ($3 - 2$) ($\nu_{\textup{rest}} = 265.88$ GHz), HCO+ ($3 - 2$) ($\nu_{\textup{rest}} = 267.56$ GHz) and HNC ($3 - 2$) ($\nu_{\textup{rest}} = 271.98$ GHz) in band 6. The channel width was 122 kHz for band 3 and 244 kHz for band 6, corresponding to velocity resolutions of 320 m s$^{-1}$ and 270 m s$^{-1}$ respectively. The total bandwidth is 468 MHz for band 3 and 937 MHz in band 6. The details of individual observations can be found in Tab.\ref{ALMAOBS}.

The Common Astronomy Software Application ({CASA} v4.7.2, \citealt{McMullin07}) was used to perform calibration and imaging. Standard calibrations were performed for atmosphere, water vapor radiometer phase corrections, bandpass, amplitude and phase calibrations following the ALMA data reduction process. Cen~A is a compact and strong continuum source that facilitates using self-calibration to improve the signal/noise. We performed phase and amplitude self calibration using line-free channels and dedicated continuum spectral windows. After the calibrations, the executions for the same schedule block were combined together for imaging using the {\sc CLEAN} algorithm with the CASA task \texttt{tclean}. The continuum source was imaged using line-free channels and continuum spectral windows using the multifrequency synthesis technique with the Briggs weighting and robust parameter +2.0 (natural weighting). The continuum emission was subtracted using the task \texttt{UVCONTSUB} using a linear fit to the line-free channels. The resulting continuum-subtracted data were then imaged with \texttt{tclean} with Briggs weighting with the same robust parameter of +2.0 to create a spectral cube. The rms noise per channel is $\sim$10~mJy. The achieved synthesized beam (angular resolution) is 2.62$\arcsec$ $\times$ 2.06$\arcsec$ and 1.4$\arcsec$ $\times$ 1.3$\arcsec$ for band 3 main array observations, 15.96$\arcsec$ $\times$ 8.49$\arcsec$ for the ACA, and 0.68$\arcsec$ $\times$ 0.62$\arcsec$ \& 1.25$\arcsec$ $\times$ 0.9$\arcsec$ for the band 6 data. The final images are presented in Figs.~\ref{fig:contour} to~\ref{fig:regions}; they were created using Astropy (\citealt{astropy}) along with Matplotlib (\citealt{matplotlib}) and are discussed in Sec.~\ref{sec:coex} below.

\subsection{{\it Chandra\/} X-ray data}
\label{data:cha}

X-ray images are needed to follow the distribution of the hot matter which may be located co-spatially with cold molecular gas. The overlapping ALMA and {\it Chandra\/} emission is a starting point in our research. Therefore, in the next step, we retrieved and analyzed a deep ($\sim$110 ksec) {\it Chandra\/} observation of Cen~A \citep[ObsID 20794, PI: P. Nulsen, see also][]{Sni19} performed on 2017 Sept 19. The observation was taken in FAINT mode, with the core placed at the aimpoint of the ACIS-S3 chip. The data were reprocessed using the \texttt{chandra$\_$repro} script of the {\it Chandra\/} Interactive Analysis of Observation (CIAO) software (v. 11) with the calibration files CALDB version 4.8.2 \citep{Fru06}. After filtering for background flares with the CIAO task \texttt{deflare}, the exposure time was $\sim$107 ksec.

An exposure-corrected image of the central region was created with the \texttt{fluximage} script. The image is in the 0.5--7 keV energy band (with an effective mean energy of 2.3 keV) and binned to 1/4 of the native ACIS pixel size (0.123$\arcsec$ pix$^{-1}$). The final image of the X-ray map from the innermost region of Cen~A combined with ALMA emissivity contours is presented in Fig.~\ref{fig:contour}, and discussed in Sec.~\ref{sec:coex} below.

\subsection{Coexistence of the hot and cold phases in Cen~A}
\label{sec:coex}

Fig.~\ref{fig:contour} shows the \textit{Chandra} X-ray extended emission (given by the colour map) with the ALMA CO line integrated flux (shown by the contours). The X-ray colour map displays emission from the AGN, the diffuse emission of the inner jet, hot gas in the central region, and point-like sources corresponding to jet knots and field sources. The hole visible at the centre of the AGN emission and the streak crossing the image in the north-west to south-east direction are artifacts (pile-up and read-out streak, respectively) caused by the brightness of the AGN.\@ The bottom image in Fig.~\ref{fig:contour} zooms in on the diffuse X-ray emission, which extends up to $5\arcsec$ ($\sim$90~pc) from the centre. The integrated CO ($1 - 0$) line flux shows the circumnuclear disk which surrounds the inner hot X-ray emission. Fig.~\ref{fig:Hlines} shows the integrated flux density map for the  HCO+ ($3 - 2$) molecular line observed in the band 6 of the ALMA main array, with the CO ($1 - 0$) line contours overlaid. Both molecular transitions are compared to our model in Sec.~\ref{sec:res}.

The ALMA data allow us to detect line luminosities from the very central region which overlaps with {\it Chandra\/} data. Since the most prominent is CO ($1 - 0$) line emission, we present the observed luminosity of this line in Tab.~\ref{tab:almareg} for different regions marked by circles in Fig.~\ref{fig:regions}. Beside the region of our interest where ALMA and {\it Chandra\/} data overlap named Reg 1, we have extracted the CO ($1 - 0$) line luminosities from neighbouring regions: Reg 2 and Reg 3, and from the most luminous region observed in the total field of view of the 7-m array, marked as Reg 4. The observed integrated flux density $S_{\rm CO}$ in Jy\,km\,s$^{-1}$ was converted to the line luminosity $L_{\rm CO}$ in erg\,s$^{-1}$ using the standard formula \citep[for details see][]{solomon97}. Luminosities from Reg 2, Reg 3, and Reg 4 are presented to demonstrate that Reg 1, the region of our interest, is the least bright. The molecular gas overlaps the hot X-ray plasma only in Reg 1, where the TI may operate. Hence, we only use Reg 1 to model the two-phase medium caused by TI in our further consideration.

\subsection{Spectral Energy Distribution}

The spectral energy distribution (SED) of the radiation that illuminates the nuclear gas in the vicinity of the SMBH is a crucial parameter in photoionization calculations. Thus it is important to use an appropriate SED corresponding to the TI mechanism operating in the source. In the case of Cen~A, which has a prominent jet, we investigate the impact of a jet-dominated radiation field on a multi-phase medium. The most adequate SED from the central region of Cen~A, presented in Fig.~\ref{cenA:sed}, was published by \citet{abdo2010} with data from different instruments, including VLBI, {\it Swift}, {\it XMM-Newton}, {\it Suzaku}, {\it Fermi\/}-LAT, and HESS, and was observed in different epochs.

The SED constructed by \citet{abdo2010} has the typical double-hump shape of non-thermal, jet-related emission. In the framework of leptonic models, the low energy curve is synchrotron emission, while the high-energy curve is usually ascribed to Compton scattering of different seed photons. In \citet{abdo2010}, two models were considered to explain the origin of the high-energy component. In the first model, the X-ray to gamma-ray band is produced via upscattering of the synchrotron photons by the electrons in the jet (synchrotron self Compton, SSC). In the second scenario, the emission was produced by upscattering of the synchrotron photons produce in a fast-moving spine by the electrons in a slow moving layer. Depending on the bulk motion of the emitting region and the jet axis inclination with respect to the observer viewing angle, the intrinsic emission can be boosted or deboosted in the observer's rest frame, and thus the overall luminosity affecting the molecular gas may vary by a factor of several, which we take into account in our computations (see Sec.~\ref{sec:param} for details). We also omit the most energetic part of the spectrum (above $10^5$~keV), which can originate from the boosted jet, and may not affect the close environment of the SMBH.\@

\subsection{Estimating the bolometric luminosity of Cen~A}
\label{data:nustar}

The bolometric luminosity of the radiation intercepting the nuclear gas is the second important input parameter for our photoionization calculations. The bolometric luminosity can be computed from the integration of the broad band observations presented in Fig.~\ref{cenA:sed}, which gives $L_{\rm bol} = 9.45 \times 10^{42}$~erg\,s$^{-1}$. But in the case of Cen~A, which is a variable source dominated by jet emission, verifying the accuracy of this value for the bolometric luminosity is challenging. Below we provide the calculations for obtaining the luminosity, variability and the contribution of the central illuminating source to the SED.\@

From observations, the 2--10 keV X-ray luminosity of the innermost hot gas of Cen~A lies in the range $5 - 9 \times 10^{41}$~erg\,s$^{-1}$ \citep{evans2004,rothschild2011}. Assuming an X-ray bolometric correction factor of $10^{1.4}$ \citep{duras2020}, the bolometric luminosity of our source from X-ray data is $L_{\rm bol} = 0.3-2 \times 10^{43}$~erg\,s$^{-1}$.
These values are compatible with the value of the bolometric luminosity obtained by integrating the SED over the solid green line marked in Fig.~\ref{cenA:sed}.

We show in Fig.~\ref{fig:lightcurve} a historical lightcurve of unabsorbed 2--10 keV flux spanning 44 years to follow the variability in the luminosity of Cen~A. Non-\textit{RXTE} data from 2000 and earlier were taken from \citet[][and individual references therein]{Risaliti2002}. We obtained \textit{Swift}-BAT data from the Swift/BAT Hard X-ray Transient Monitor website\footnote{\url{https://swift.gsfc.nasa.gov/results/transients/}} and rebinned them to 20 days, and extrapolated from the $15 - 50$ keV count rates (hereafter CR$_{15-50}$ in units of cts cm$^{-2}$ s$^{-1}$), to estimate unabsorbed $2 - 10$ keV power-law fluxes. Based on the 70-month average BAT spectrum of Cen~A, we assumed a power-law photon index of 1.87, and a conversion of $F_{2-10}$ ($10^{-11}$ erg cm$^{-2}$ s$^{-1}$) = $5726$~CR$_{15-50}$. Reduction of \textit{Suzaku} XIS and HXD data followed standard extraction procedures, and using spectral models based on \citet{Markowitz2007} and \citet{Fukazawa2011}. The 2013 Nuclear Spectroscopic Telescope ARray (\textit{NuSTAR}) flux was estimated from the best-fit spectrum of \citet{furst16}. Reduction and fitting of the 2015 \textit{NuSTAR} data followed \citet{furst16} and will be detailed in a future paper by Markowitz et al., (in prep.).
Reduction and fitting of the 2018 and 2019 \textit{NuSTAR} data also followed \citet{furst16} and is discussed in further detail for the 2018 observation below. Reduction and fitting of \textit{XMM-Newton} EPIC pn data followed \citet{evans2004}. Reduction and fitting of \textit{RXTE}-PCA data followed standard procedures, e.g., \citet{Rivers2011}. Reduction and fitting of \textit{AstroSAT} SXT and LAXPC data will be detailed in a future paper (Markowitz et al., in prep.). Assuming a luminosity distance of 3.8 Mpc \citep{rejkuba2004}, this flux yields an (assumed isotropic) 2--10 keV luminosity of $ 1.1 \times 10^{42}$ erg s$^{-1}$ (including BAT;\@ $ 1.2 \times 10^{42}$ erg s$^{-1}$ excluding BAT). Fig.~\ref{fig:lightcurve} clearly demonstrates that the X-ray flux can change by a factor of roughly 10 over timescales of $\sim$ a decade.

To derive a robust AGN X-ray luminosity and jet contribution for Cen~A (2--10 keV band), we used the \textit{NuSTAR} observations \citep{Harrison13}. \textit{NuSTAR} encompasses the 3--78 keV energy band and has a broader point spread function (PSF) than \textit{Chandra}, with a half-power diameter of $60\arcsec$ \citep{madsen2015}. The instrument uses triggered readout to log detected photon events that eliminates pile-up. The nearest \textit{NuSTAR} observation in time to the \textit{Chandra} observation analysed here is ObsID 60466005002, performed on 2018 Apr 23. We produced calibrated and cleaned events files and extracted spectra from $100\arcsec$ circular regions for each focal plane module (FPM). We then extracted background spectra from a circular region as large as possible on the same detector chip as the source within each FPM.\@

We fit the \textit{NuSTAR} spectrum extracted for the AGN, with each FPM spectrum binned to a minimum of 25 counts per bin. Our model is an absorbed power-law and Gaussian Fe\,K$\alpha$ emission line. This gives $\chi^{2}$\,/\,$n$\,=\,1344.88\,/\,1381, where $n$ is the number of degrees of freedom. The total observed flux for this model is
$F_{\rm 2-10\,keV}$\,=\,2.4\,$\times$\,10$^{-10}$\,erg\,s$^{-1}$\,cm$^{-2}$
in both FPMA and FPMB.\@ At the distance of Centaurus~A ($D$\,=\,3.8\,Mpc), this gives an observed (i.e.\ absorbed) luminosity of log\,$L_{\rm 2-10\,keV}$\,/\,erg\,${\rm s}^{-1}$\,=\,41.6 for the AGN.\@ This corresponds to an unabsorbed luminosity of log\,$L_{\rm   2-10\,keV}$\,/\,erg\,${\rm s}^{-1}$\,=\,41.9 for the AGN in both FPMA and FPMB.\@

To test the contribution of contaminants in the \textit{NuSTAR} spectrum, we manually extracted off-nuclear sources of flux in the \textit{Chandra} events file for Centaurus~A. A total of 10 off-nuclear point-source contaminants and both the jet and counter jet components were clearly visible with \textit{Chandra} located inside the extraction region used for the \textit{NuSTAR} source spectrum. By fitting each contaminant separately, and summing the predicted 2--10\,keV flux, we find that the observed 2--10\,keV flux for the AGN is a factor of $\sim$50 times higher than the combined flux from the 12 total contaminants manually extracted, further indicating that we are estimating the AGN flux robustly with \textit{NuSTAR}.

This gives a 2--10\,keV luminosity of $L_{\rm X} = 3.98 \times 10^{42}$~erg\,s$^{-1}$ for the AGN in Cen~A. Taking into account the bolometic correction factor, $L_{\rm bol} = 1 \times 10^{44}$~erg\,s$^{-1}$ at the date closest to the {\it Chandra\/} observations. Assuming that only ten percent of this emission originates from the inner hot radius, while the rest comes from the jet, we take $L_{\rm bol} = 1 \times 10^{43}$~erg\,s$^{-1}$ as a reference value for our calculations.

After considering the information on the uncertainty of the jet beaming factor, the variability of X-ray flux, and the bolometric luminosity estimation, we followed our first paper \citep{rozanska14} where several luminosity states for Sgr~A* were considered in photoionization calculations. In the case of Cen~A studied here, we consider two luminosity states, differing by a factor of 70, and keeping the same overall SED shape as given in Fig.~\ref{cenA:sed}. When Cen~A has normal high luminosity state we assume our reference value of $L_{\rm bol} = 1 \times 10^{43}$~erg\,s$^{-1}$, while for the low luminosity state we take $L_{\rm bol} = 1.43 \times 10^{41}$~erg\,s$^{-1}$. All important properties of the Cen~A inner region used for our simulations are listed in Sec.~\ref{sec:param}.

\section{Model set-up}
\label{sec:model}

To model the multi-phase gas in the centre of Cen~A, we use the same approach as in our previous work dedicated to Sgr~A* \citep{rozanska14} and M60-UCD1 \citep{rozanska17}. The irradiation-induced effects of TI \citep{field1965} operate only for a certain range of parameters \citep[e.g.,][and further references therein]{cox2005}. The most important parameter is the occurrence of strong ionisation, which means that (I) interstellar medium (ISM) gas should be metal rich \citep{hess1997}, and (II) an illumination field should be hard enough to be able to remove all electrons even from heavy atoms, and display substantial luminosity. Such radiation is often observed from central accreting SMBHs, and can provide support for a multi-phase medium in the close vicinity of a SMBH \citep{krolik1981,rozanska96}.

The radiation field may originate from the extended inner part of the jet, nuclear star cluster (NSC) stars or from the hot plasma around the SMBH.\@ In our previous work concerning Sgr~A* and M60-UCD1 \citep{rozanska17}, we have taken into account the additional radiation field from the NSC, but in the case of Cen~A, the starburst activity is negligible compared to the emission from the AGN \citep{Radomski08}. Therefore, for the purpose of this paper we neglect any additional energy input by photons from stars.

Further complications may arise when the region under consideration is dominated by mechanical energy input from stellar winds or outflows \citep{quataert2004,silich2004,silich2008}. This was accounted for the case of Sgr~A* and M60-UCD1, where such energy input by winds was reported \citep[see discussion in][]{rozanska17}. In the case of Cen~A it has been reported that close to the SMBH, the X-ray dominated region (XDR) dominates over mechanical heating \citep{furst16, espada17}. Under these circumstances, we neglect mechanical energy input in our computations. For the full treatment of the multi-phase media, hydrodynamic simulations with radiation should be made \citep[see][and references therein]{waters2019, Dannen2020}, but those simulations still do not consider many atomic processes i.e.\ lines, and cannot be directly compared with data.

Depending on the profile of the SED, the incident radiation field can both cool and heat the plasma. The net effect is determined by plasma density as well as the spectral shape of the incident radiation. Thus, the presence of the radiation field contributes both to the cooling and heating rates, ${\cal L}$ and ${\cal H}$, respectively, in erg s$^{-1}$ cm$^{-3}$. The difference between these two parameters describes the rate of exchange of total energy per unit volume by different cooling and heating processes.

Another essential parameter of our model is the gas structure. Recent studies \citep{rozanska2006,Baskin2014,Adhikari2015,Adhikari2018,Adhikari2019,Adhikari2019A} have demonstrated that the gas clouds illuminated by the radiation field in various astrophysical environment remain in total (radiation + gas) pressure equilibrium, as it is dominated by the radiation pressure. We refer the reader to \citet{Adhikari2018} and \citet{Adhikari2019} for the relevant discussion on the validity and differences between the constant pressure and constant density assumptions used in the photoionization modelling. Following this, we adopt the assumption of constant pressure in all the model commutations presented in this paper.

To solve for the stability of thermal equilibrium, we employ {\sc cloudy}~17.01 \citep{Ferland2017} which computes a transfer of radiation through matter with different geometries, allowing us to chose an ISM abundance including dust. All important radiative processes which heat or cool plasma are taken into account, such as free-free emission, atomic bound-free processes, and bound-bound transitions between different ion states. This solution provides us with the local equilibrium temperature of the medium as a function of assumed ISM hydrogen density at a given location. The stability can be conveniently estimated from the stability curve. The definition of the ionisation parameter $\Xi$ \citep{tarter69} is
\begin{equation}
\Xi = \frac{ L}{ 4 \pi c P_{\rm gas} R^2},
\label{eq:bigxi}
\end{equation}
where $L$ is the source luminosity, $P_{\rm gas}$ is the local gas pressure, and $R$ is the distance from the centre. The $\Xi$ vs.\ $T$ curve
indicates the thermal stability. In this relation, the branch with the positive slope is radiatively stable while the branch with the negative slope is unstable.

The equation above requires the knowledge of additional parameters in order to construct a full photoionization model. Besides the SED and bolometric luminosity, we have to assume the density of the matter at the illuminated cloud surface, $n_{\rm H}$, the total column density of the nuclear gas $N_{\rm H}$,
and the location of the matter from the nucleus $R$. In case of Cen~A, the location of cold and hot material can be adopted from the high-resolution images presented above.

\subsection{Model parameters for Cen~A}
\label{sec:param}

For the purpose of this paper we adopt the averaged value of $M_{\rm BH}=6 \times 10^7$~M$_{\odot}$ for the black hole mass. For such a
mass, the gravitational radius is $R_{\rm g} = 8.12 \times 10^{12}$~cm, and the Schwarzschild radius is $R_{\rm Schw} = 1.62 \times 10^{13}$~cm. We assume a cloud at a certain distance from the nucleus with abundances that include heavy elements as is typical for Solar abundances. Alternatively, we also consider a cloud consisting of a mixture of gas and dust in the proportions appropriate for the ISM (a standard option in {\sc cloudy}).

For all cases, we consider the shape of the illuminated SED as presented in Fig.~\ref{cenA:sed} by the green solid line. We consider two luminosity states discussed in Sec.~\ref{sec:mainp} with the same spectral shape. When Cen~A is in the high luminosity state we assume $L_{\rm bol} = 1 \times 10^{43}$~erg\,s$^{-1}$, and for the low luminosity state we take $L_{\rm bol} = 1.43 \times 10^{41}$~erg\,s$^{-1}$.

We assume hot plasma filling the space with density $n(r) = n_0*{(r/r_0)}^{-1.5}$. From older {\it Chandra\/} observations, \citet{kraft2003} estimated the density $n_0 = 3.7 \times 10^{-2}$~cm$^{-3}$, and $r_0 = 10$~pc, which we use for our reference model of a single cloud with density $n_{\rm H_{0}} = 3.7 \times 10^{-2}$~cm$^{-3}$ at the cloud surface, located at 10 pc from the nucleus and illuminated by the high luminosity state. We also use additional values of density to test the dependence of results on density (Sec.~\ref{sec:res}). We assume that each cloud is under constant pressure, and thus the value of density is defined only at the illuminated cloud surface. Under this assumption, the density and ionization structure is stratified deep inside the cloud if a large enough column of matter is considered.

We construct the stability curve using {\sc cloudy} simulations to test the stability of gas clouds. We conducted the stability analysis in two ways. (I) To reconstruct the stability curve at a certain radius, we adopt geometrically-thin clouds with densities spanning $\log(n_{\rm H_{0}}[\rm cm^{-3}])$ from  $-3$ to $13$. (II) Secondly, a low-density, geometrically-thick single cloud illuminated by the incident SED can be used to reconstruct the stability curve. Such a cloud, being under constant pressure, becomes strongly stratified due to irradiation.

When constructing the stability curve for thin clouds of different densities, we perform photoionization calculations for three distinct distances from the nucleus at which the clouds are located. In other words, we construct three stability curves at distances:
R = 10$\arcsec$ (180~pc),
R = 3$\arcsec$ (55~pc),
and R = 0.62$\arcsec$ (11.5~pc).
The last value is the best spatial resolution of ALMA in our observations. The location of the single low-density, geometrically-thick cloud used in the photoionization modeling depends on the location of the hot illuminated cloud surface, and signifies how far from the nucleus cold matter is extended. We place the illuminated cloud surface at three distances:
R = 2.71$\arcsec$ (50~pc),
R = 1.09$\arcsec$ (20~pc), and R = 0.54$\arcsec$ (10~pc).
The smallest 10~pc radius is not resolved in the X-ray observations, but it is used to calculate the temperature of the inner plasma. All the distances are still significantly larger than the gravitational radius.

\begin{figure*}
  \includegraphics[width=15cm]{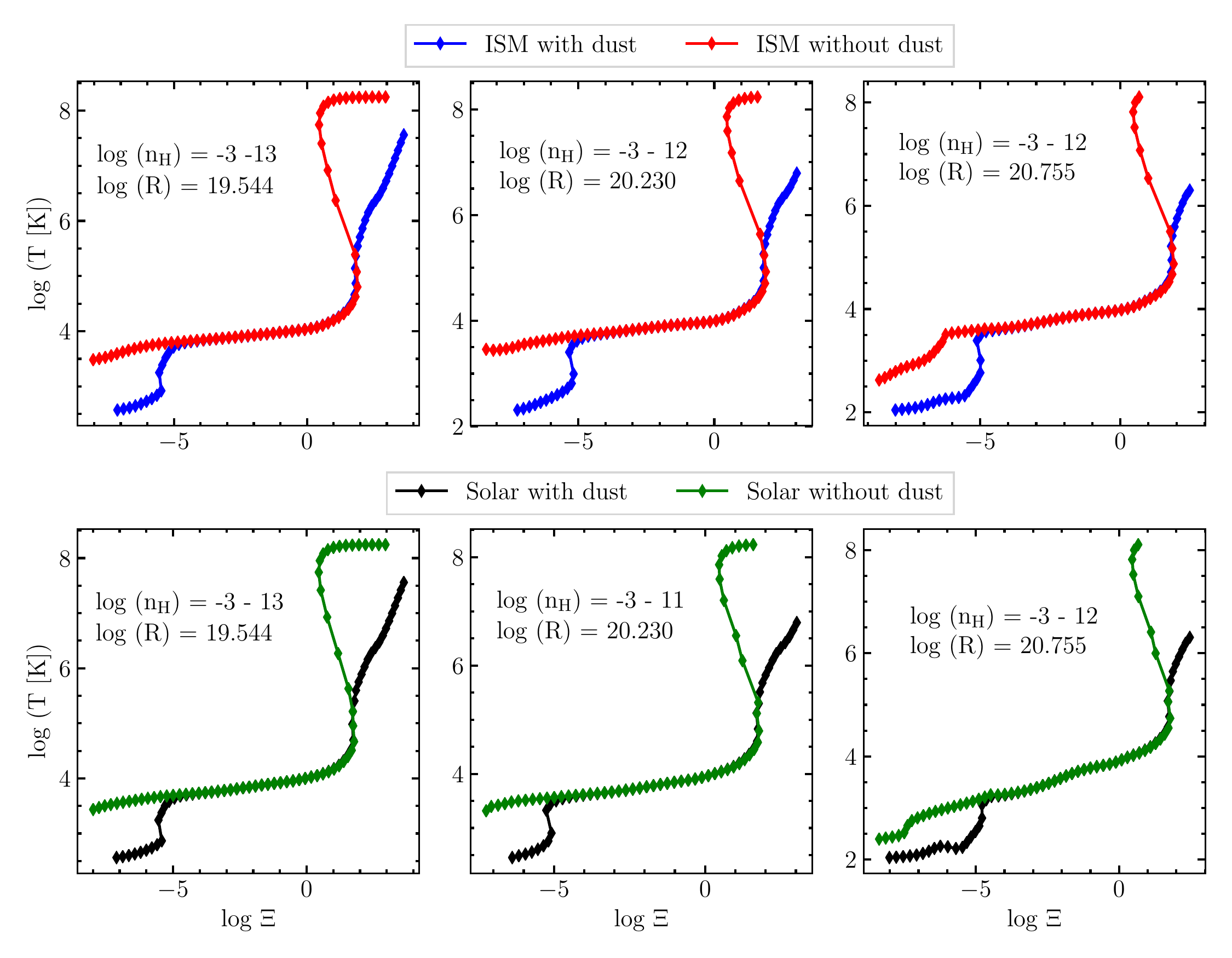}
  \caption{Stability curves for two different heavy element abundances: upper panels show ISM abundances, while bottom panels show Solar abundances, with increasing radii from left to right: 11.5, 55, and 180 pc. Each panel contains calculations with and without dust. Each point in the curve represents the model computed for one starting value of density. The range of computed densities is given in each panel. The SED luminosity is always kept at $L_{\rm bol}=1 \times 10^{43}$~erg\,s$^{-1}$.}
\label{fig:stab}
\end{figure*}

\section{Results}
\label{sec:res}
The stability curves for three different distances from the centre and for two different chemical abundances are presented in Fig.~\ref{fig:stab}. The upper panels of the figure display the curves computed for the ISM abundances with dust (blue points) and without dust (red points) while the lower panels display the curves for Solar composition with dust (black points) and without dust (green points). Each point represents a thin cloud under constant pressure and illuminated by the same radiation field of the Cen~A SED and reference value of $L_{\rm bol}$. As the ionization parameter $\Xi = P_{\rm rad}/P_{\rm gas}$, vertical lines on the stability curves indicate clouds under constant pressure. Figure~\ref{fig:stab} shows that hot gas with temperature of the order of $10^8$~K emitting in X-rays can coexist with cold matter at $\sim 10^4$~K emitting in visible light. Nevertheless, dusty clouds do not coexist with hot X-ray plasma, and can only coexist with cold gas, due to evaporation from X-rays. Note that changing the ISM composition to Solar only slightly influences the shape of the stability curve and only in cold dusty regions. Since cold regions are most probably composed of interstellar matter, we use ISM composition for further considerations.

\begin{figure*}
  \includegraphics[width=16cm]{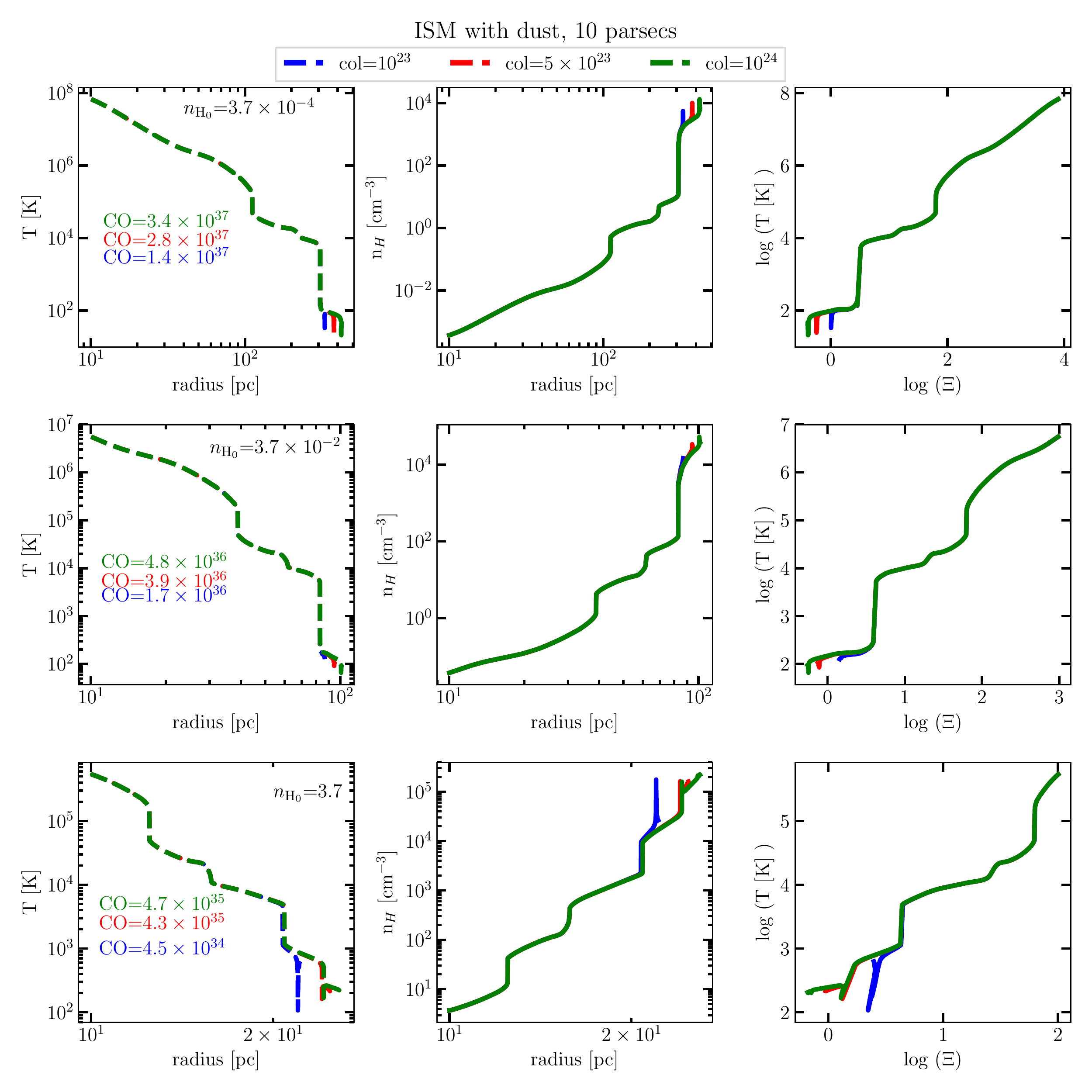}
  \caption{A single geometrically-thick cloud structure an illuminated cloud surface located at 10~pc for three different values of starting density: $3.7 \times 10^{-4}$ cm$^{-3}$ in the upper panels, $3.7 \times 10^{-2}$ cm$^{-3}$ \citep[see][]{kraft2003} in the middle panels, and 3.7~cm$^{-3}$ in the lower panels. From left to right the three columns represent temperature radial structure in the left column, density radial structure in the middle column (plotted over the depth of the cloud in both columns), and the stability curves (plotted over the ionization parameter) in the rightmost column. The models are computed for a radiation luminosity of $10^{43}$ erg~s$^{-1}$. The colors of the solid lines indicate different total column densities assumed in the computations, as shown on top of the figure; color labels give the predicted CO ($1 - 0$) luminosities (in erg s$^{-1}$) corresponding to the adopted column density.}
\label{fig:lane10}
\end{figure*}

\begin{figure*}
  \includegraphics[width=16cm]{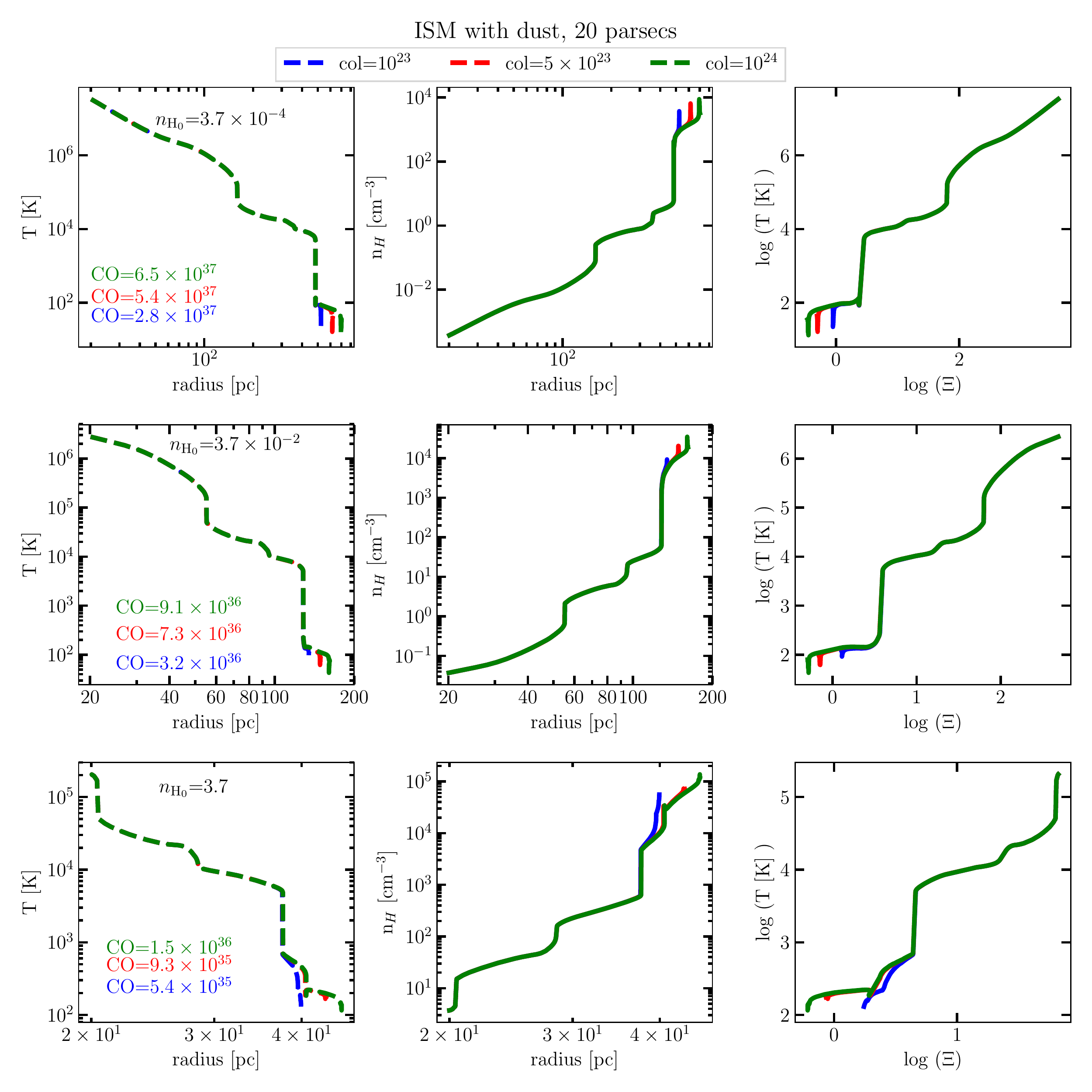}
  \caption{Same as in Fig.~\ref{fig:lane10}, but for an inner radius of 20~pc.}
\label{fig:lane20}
\end{figure*}

\begin{figure*}
  \includegraphics[width=16cm]{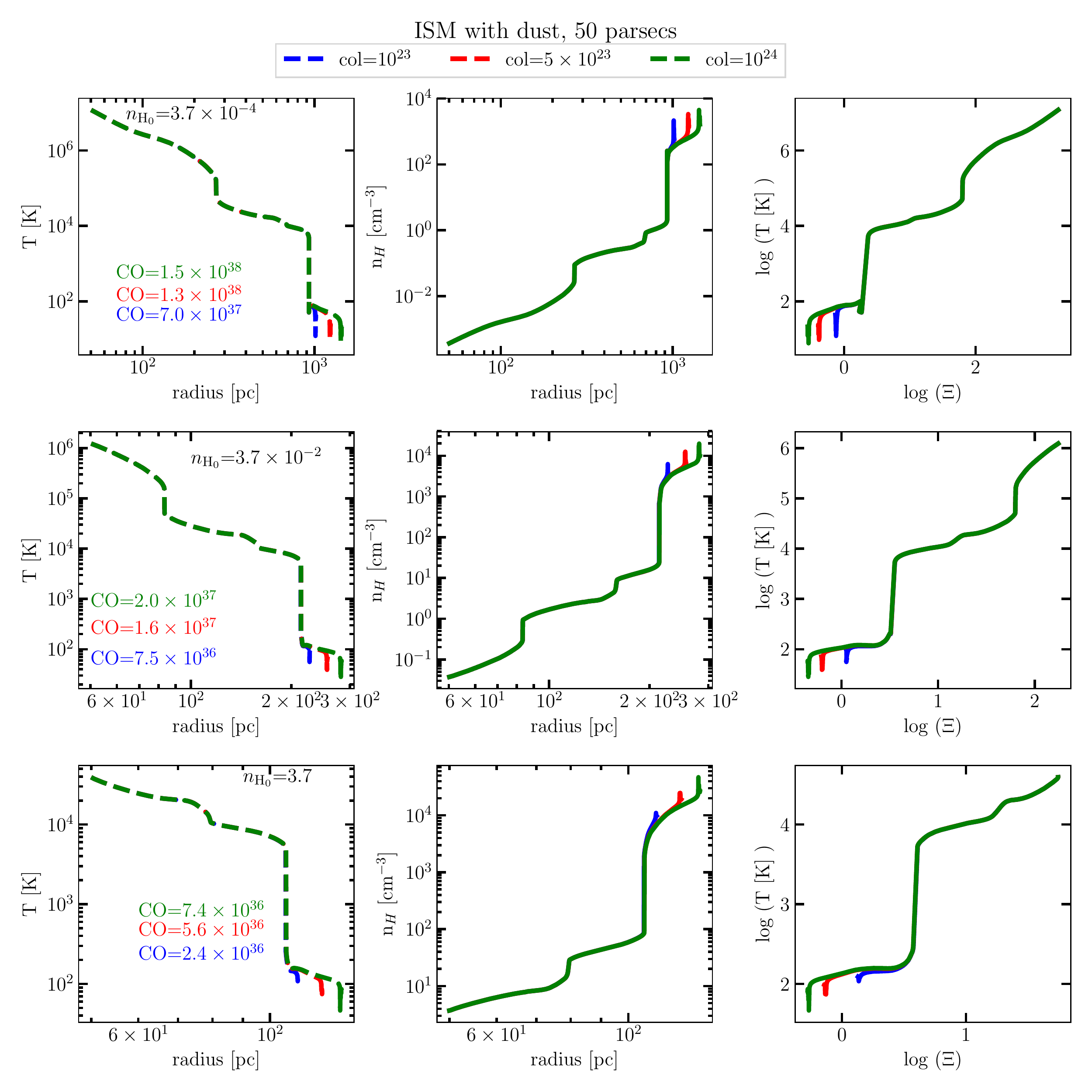}
  \caption{Same as in Fig.~\ref{fig:lane10}, but for an inner radius of 50~pc.}
\label{fig:lane50}
\end{figure*}

Next, we computed various model structures for geometrically thick dusty clouds, located at representative distances of 10, 20 and 50~pc, and for relatively large column densities from $N_{\rm H}=1 \times 10^{23}$ up to $1 \times 10^{24}$~cm$^{-2}$. The large values of $N_{\rm H}$ ensure that the radiation passes across the depth of cloud and reaches the minimum temperature at which the molecular emission is significant. The temperature, density structures and stability curves for radii 10, 20, 50 pc are presented in Figs.~\ref{fig:lane10},~\ref{fig:lane20}, and~\ref{fig:lane50} respectively. These clouds are computed for an open thick geometry, with the assumption of constant pressure, and for the following values for the hydrogen number density at the surface of the cloud:
$n_{\rm
  H_{0}}=3.7 \times 10^{-4}$~cm$^{-3}$ (upper panels), $n_{\rm
  H_{0}}=3.7 \times 10^{-2}$~cm$^{-3}$ from \citet{kraft2003} (middle panels), and $n_{\rm H_{0}}=3.7$~cm$^{-3}$ (lower panels) in each of these figures.

The temperature profiles presented in the right panels of Figs:~\ref{fig:lane10},~\ref{fig:lane20}, and~\ref{fig:lane50} clearly indicate a multi-phase medium. They display jumps at the instability zones. For the least dense cloud considered ($n_{\rm H_{0}}=3.7 \times 10^{-4}$~cm$^{-3}$), the material extends up to $\sim$ 430~pc ($23\arcsec$), but in the case of the higher starting density ($n_{\rm H_{0}}=3.7$) the material extends only up to 20~pc ($1\farcs1$). The hot part indeed may be spherical, but changing the geometry in {\sc cloudy} from open thick to closed i.e.~spherical, does not change our results considerably. The CO line emission slightly decreases when the open geometry is replaced by the spherical geometry (see {\sc cloudy} documentations for details).

We present the luminosities of several molecular lines computed by {\sc cloudy} for two individual clouds of different values of $n_{\rm H_{0}}$ in Tab.~\ref{tab:almalines} (sixth and seventh columns). The same lines are observed by ALMA and the measured luminosities for Reg 1, following the standard formula \citep{solomon97}, are listed in the third column of Tab.~\ref{tab:almalines}. The $^{13}$CO~($1 - 0$) line is not represented in our models. All other lines can be directly compared with observations but it is not trivial to find one model which reproduces luminosities of all lines with the same parameter configuration. Since the CO line is the most prominent in the ALMA data, our modeling is focused on the CO line.

\begin{table*}
  \begin{center}
    \caption{\small The most intense line luminosities measured by ALMA from Reg 1, shown by red circle in the images given in Fig.~\ref{fig:regions}. The first two columns display the line transition and its frequency, while the line luminosities integrated over a circular surface with radius $4\farcs5$ are given in column 3. The critical densities above which lines are visible are displayed in column 4. The corresponding wavelength of each line is given in column 5. The line luminosities obtained from {\sc cloudy} computations for two cloud locations, 10 and 50 pc, are listed in columns 6 and 7 respectively.}
  \begin{tabular}{lrrrrll}
  \hline
  \multicolumn{3}{c}{Observations} & \multicolumn{4}{c}{Modelling} \\
  \hline
  Line & Freq. ALMA & Lum. ALMA & $n_{\rm crit}$ & Wavelength & Lum. {\sc cloudy} & Lum. {\sc cloudy}\\
  Name &[GHz] & [erg\,s$^{-1}$] & [cm$^{-3}]$ & [$\mu$m] & [erg\,s$^{-1}$] & [erg\,s$^{-1}$]\\
  & & & & & $n_{\rm H_{0}}=3.7\times10^{-4} $~cm$^{-3}$ & $n_{\rm H_{0}}=3.7 \times10^{-4}$~cm$^{-3}$ \\
  & & & & & $r_{\rm cl} = 10-420$~pc & $r_{\rm cl} = 50-1400$~pc \\
  \hline
  CO ($1 - 0$) & 115.271202 & $ 1.97 \times 10^{38}$ & $1 \times 10^{3}$ & 2600.05 & $3.37 \times 10^{37}$ & $1.51\times 10^{38}$ \\
  $^{13}$CO ($1 - 0$) & 110.201354 & $1.97 \times 10^{37}$ & & & & \\
  CS ($2 - 1$) & 97.980953 & $1.69 \times 10^{37}$ & $1.3 \times 10^{5}$ & 3058.86 & $2.43 \times 10^{37}$ & $9.34\times 10^{37}$ \\
  HCN ($3 - 2$) & 265.886431& $9.87 \times 10^{37}$ & $1.4 \times 10^{7}$ & 1127.22 & $1.35 \times 10^{38}$ & $5.89\times 10^{38}$ \\
  HCO+ ($3 - 2$) & 267.557633 & $1.06 \times 10^{38}$ & $1.6 \times 10^{6}$ & 1120.18 & $3.13 \times 10^{37}$ & $6.37\times 10^{37}$\\
  HNC ($3 - 2$) & 271.981111 & $6.24 \times 10^{37}$ & $5.1 \times 10^{6}$ & 1101.95 & $2.52 \times 10^{38}$ & $ 9.41\times 10^{38}$\\
\hline
\end{tabular}
\label{tab:almalines}
\end{center}
\end{table*}

\begin{figure}
  \includegraphics[scale=0.45]{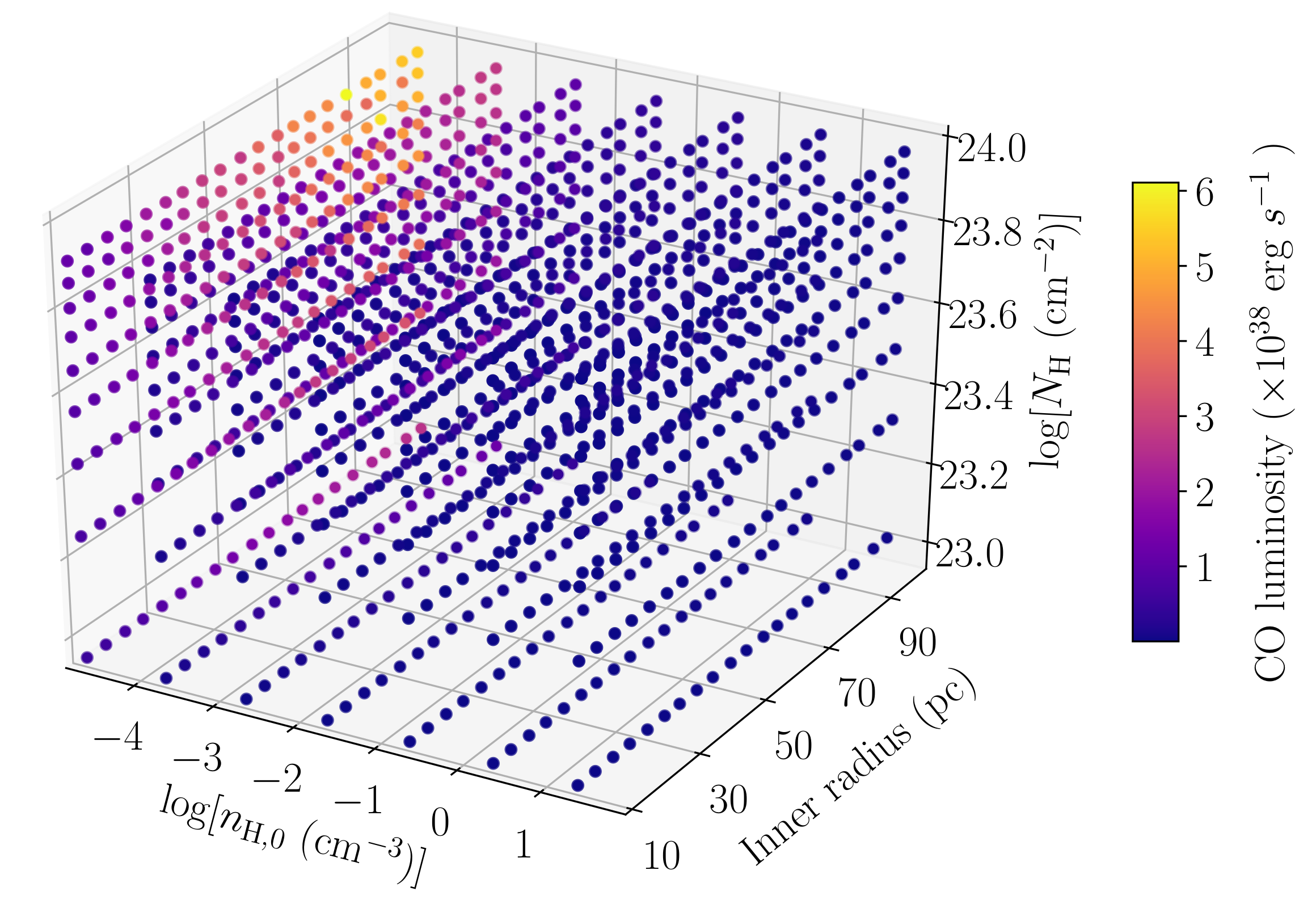}
\caption{3-D plot of CO ($1 - 0$) line emission computed in {\sc cloudy} as a function of gas density, inner radius, and
the column density.}
\label{fig:3dco}
\end{figure}

\begin{figure}
  \includegraphics[scale=0.45]{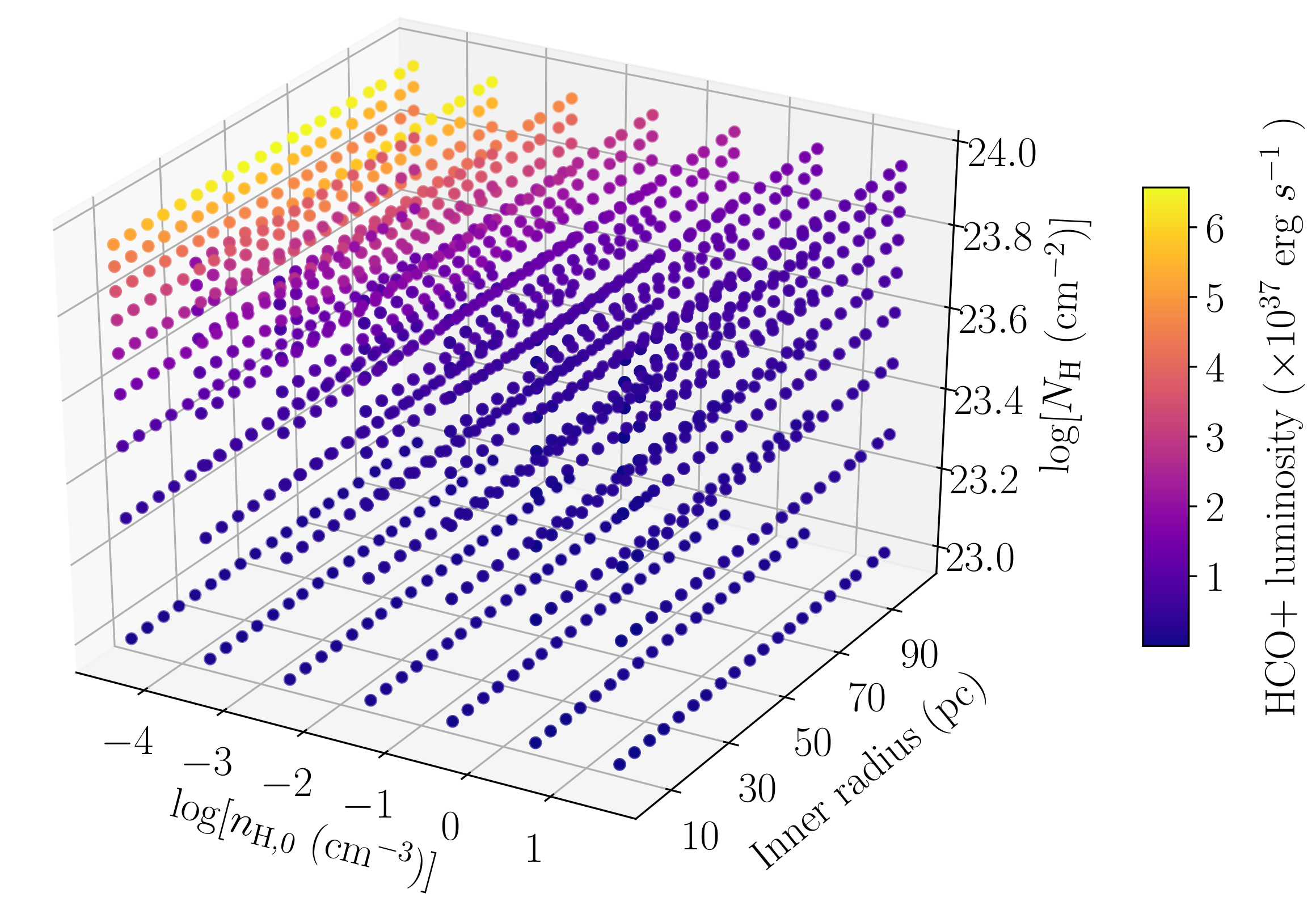}
\caption{3-D plot of HCO+ ($3 - 2$) line emission computed in {\sc cloudy} as a function of gas density, inner radius, and
the column density.}
\label{fig:3dhco}
\end{figure}

\subsection{Dependence on density}

It is well known that there is a critical density for the thermal molecular emission, i.e.\ density above which the particular line emission is visible at a certain luminosity \citep{loenen07, shirley15}. The list of critical densities for the lines observed by ALMA in Cen~A, integrated over a circle of radius $4\farcs5$, is given in Tab.~\ref{tab:almalines} in the fourth column.

To study the density dependence, we searched a large parameter space with {\sc cloudy}, using a single cloud under constant pressure for each set of input parameters, to see which clouds closely reproduce the molecular line luminosities deduced from ALMA data as presented in column 3 of Tab.~\ref{tab:almalines}. Figs~\ref{fig:3dco} and~\ref{fig:3dhco} depict the gas densities, column densities and radii used, and the resulting molecular emission obtained from {\sc cloudy} models for the lines CO~($1 - 0$) and HCO+ ($3 - 2$) respectively.

From Figs.~\ref{fig:3dco} and~\ref{fig:3dhco}, it is quite clear that the molecular line emission strongly depends on the value of the gas density used in the photoionization models. A slight variation in line luminosities with radial distance is also seen from both figures. In our case, enhanced molecular emission is obtained for the model with the lowest value of gas density considered at the illuminated face of the cloud i.e., $3.7 \times 10^{-4}$ cm$^{-3}$. This is due to the fact that the CO emission area in a low-density cloud is much larger than in the case of a high-density cloud. The difference between these two cases can be realized by comparing the first and third rows of Fig.~\ref{fig:lane10}. When the density increases by 4 orders of magnitude, the CO emission is reduced by 2 orders of magnitude. Note that this reduction is directly related to the size of the emitting cloud before it reaches the backside of the cloud. The size of the CO emission cloud for the least dense cloud ($3.7 \times 10^{-4}$ cm$^{-3}$) is $\sim$ $400$ pc, whereas the size for the densest cloud ($3.7$ cm$^{-3}$) is $\sim$ $28$~pc (Fig.~\ref{fig:lane10}).

\begin{figure}
  \includegraphics[width=8.5cm]{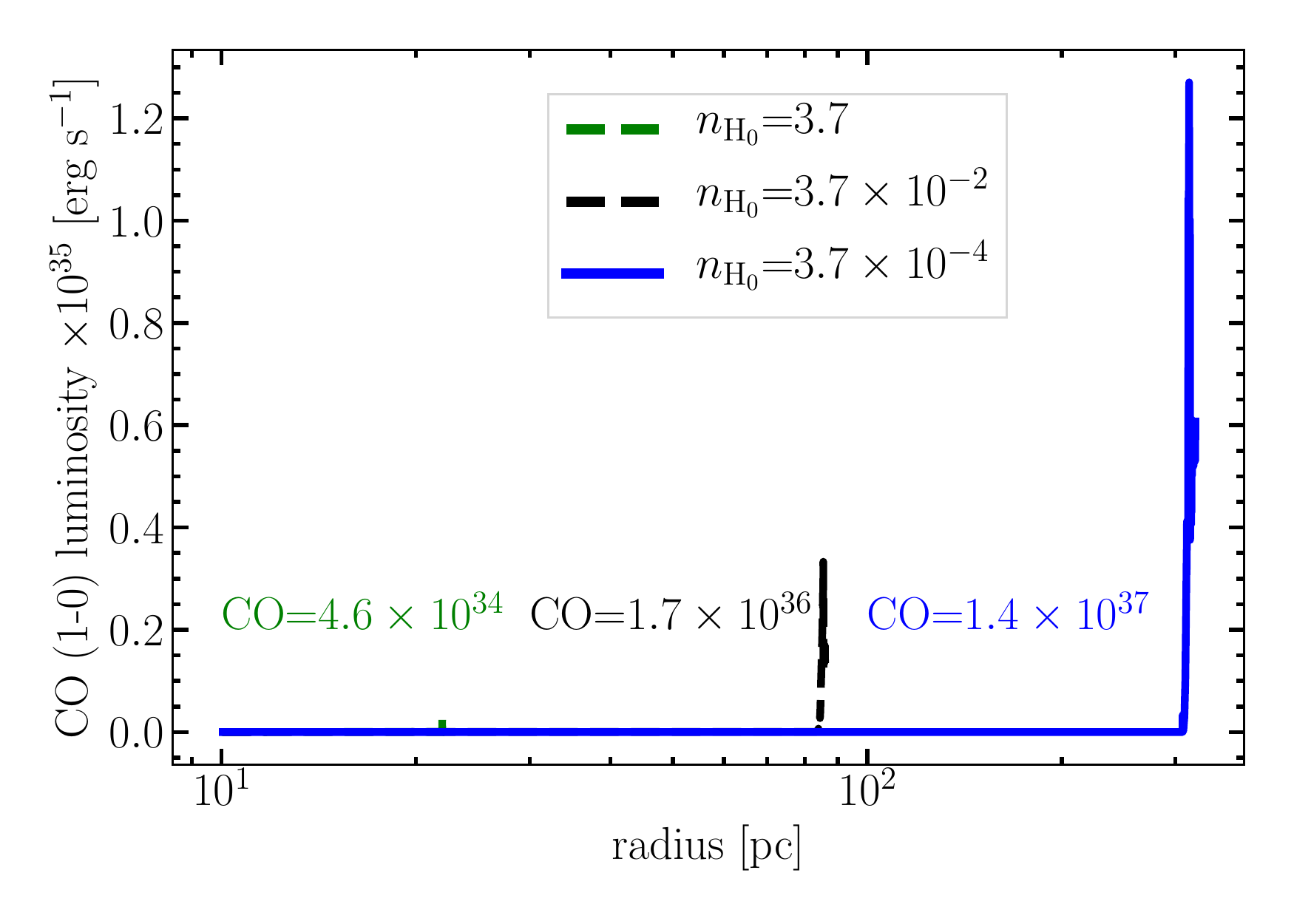}
  \caption{CO ($1 - 0$) luminosity as a function of radial distance. Emission for three cases of initial densities, $n_{\rm H_{0}}$= $3.7 \times 10^{-4}$, $3.7 \times 10^{-2}$ and $3.7$ cm$^{-3}$, are shown. The total luminosity is obtained by integrating the luminosity over the whole range of radii.}
\label{fig:emissivity}
\end{figure}

Fig.~\ref{fig:emissivity} shows the CO~($1 - 0$) luminosity as a function of the radial distance across the cloud. A comparison between the two initial density cases $3.7 \times 10^{-4}$ and $3.7$ cm$^{-3}$ demonstrates that the CO emission comes from a large distance for the less dense gas. However, most of the emission is produced in the thin layer of the cloud as compared to its total size. This is due to the fact that CO~($1 - 0$) is produced in the low temperature region close to the backside of the illuminated cloud.

\subsection{Dependence on the source luminosity}

\begin{figure*}
  \includegraphics[width=16cm]{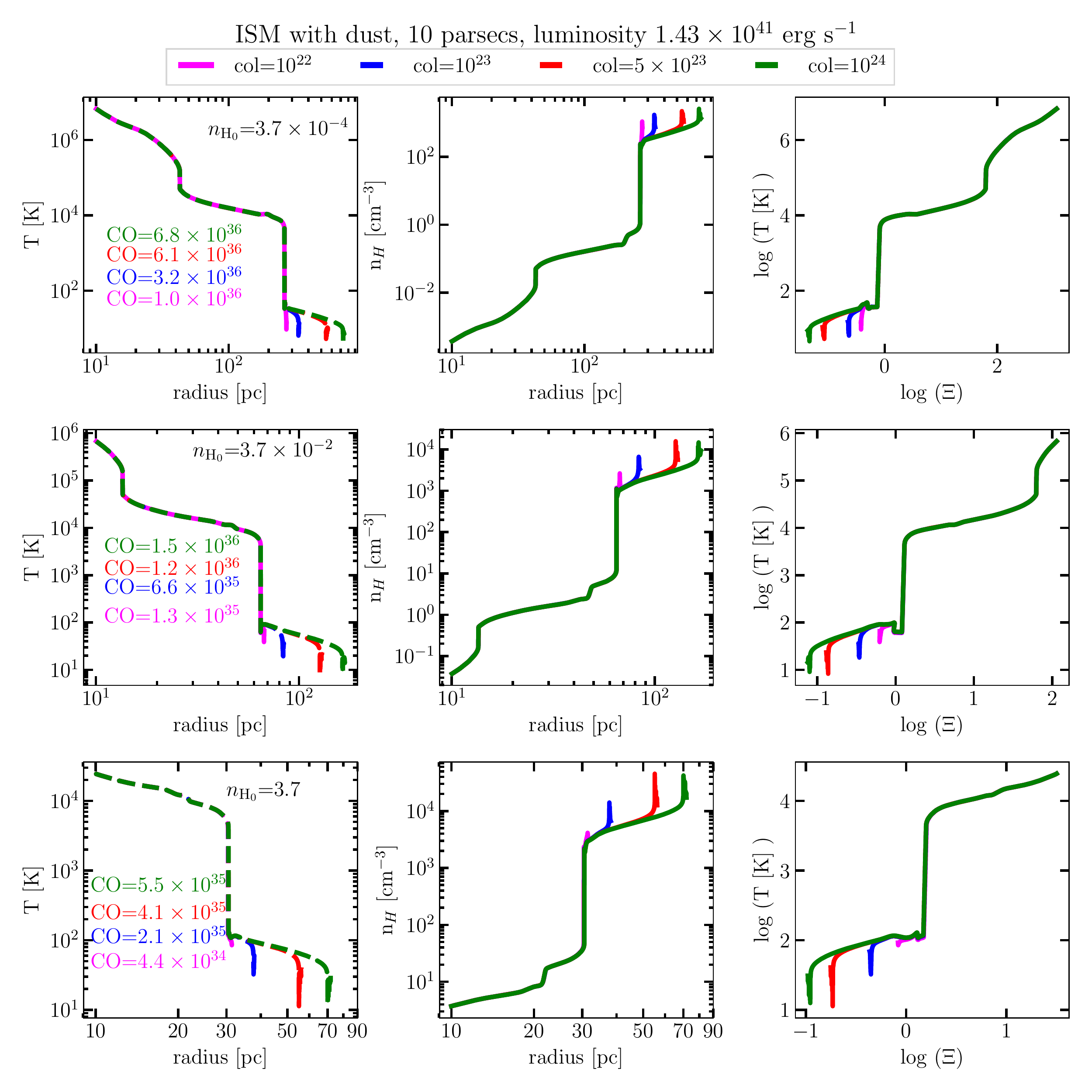}
  \caption{The same as in Fig.~\ref{fig:lane10} but for the lower value of radiation luminosity, $1.43 \times 10^{41}$ erg s$^{-1}$.}
\label{fig:lane50_lum41}
\end{figure*}

In order to study the effect of the luminosity on the molecular emission, we adopted the luminosity value $1.43 \times 10^{41}$ erg s$^{-1}$, which is lower by a factor of $\sim$ 70 than the value considered previously, and we computed a {\sc cloudy} single thick cloud model under the constant pressure condition at $10$ pc radius. Only the source luminosity was changed while keeping the SED same as in Fig.~\ref{cenA:sed}. For such a low luminosity, the temperature, density and ionization structure computed for three different densities considered is shown in Fig.~\ref{fig:lane50_lum41}. Comparison between Figs.~\ref{fig:lane10} and~\ref{fig:lane50_lum41} demonstrates that the overall structure of the cloud does not differ significantly. However, the CO~($1 - 0$) luminosity is an order of magnitude lower for low density cloud for the low luminosity case (upper panels). Nevertheless, the difference in line emissivity for two luminosity cases decreases with the increasing cloud density (towards lower panels).

\begin{figure*}
  \includegraphics[width=11cm]{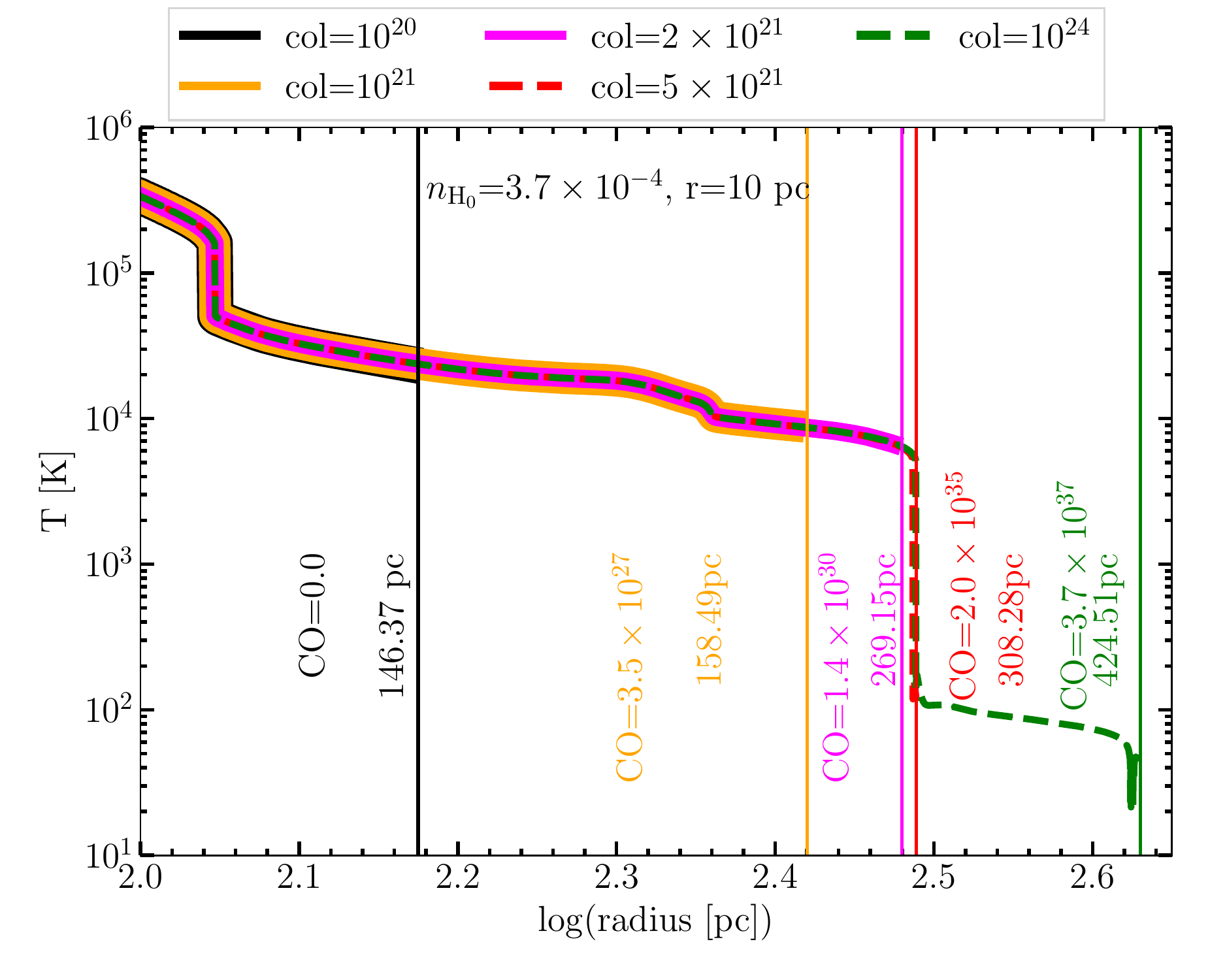}
  \caption{Length scales of CO ($1 - 0$) emission for various column densities are given in the panel box. The model is computed for a gas density $3.7 \times 10^{-4}$ cm$^{-3}$ located at the inner radius of 10~pc. For each case the cloud outer radius is marked by a vertical line. The resulting cloud sizes and CO~($1 - 0$) line luminosities are displayed near each vertical line. Units of luminosity are erg s$^{-1}$, and units for column density are cm$^{-2}$.}
\label{fig:rad_10pc_scales}
\end{figure*}

\begin{figure*}
  \includegraphics[width=11cm]{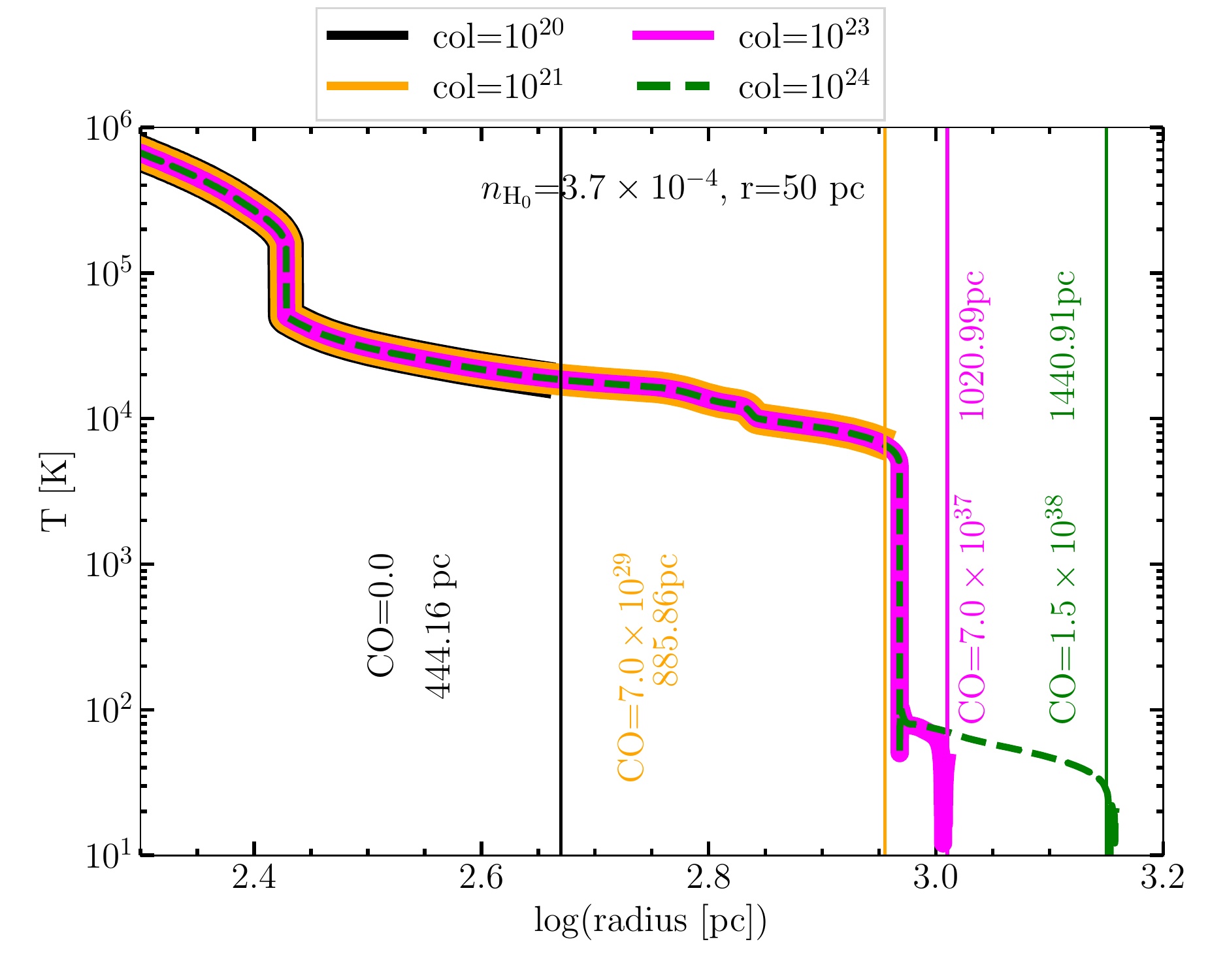}
  \caption{Same as Fig.~\ref{fig:rad_10pc_scales}, but for r $=50$~pc.}
\label{fig:rad_50pc_scales}
\end{figure*}

\subsection{Dependence on cloud column density}

For the single thick cloud consideration, the increase of cloud column density implies that more cold matter is added on the back, non-illuminated side of the cloud. Therefore, the line emissivity increases with the column density, but only when the starting gas density is relatively low, as seen in Figs.~\ref{fig:3dco} and~\ref{fig:3dhco}. It is noteworthy that the column density values where CO and HCO+ reach their maxima agree with recent estimates of radio and X-ray column densities for compact radio galaxies \citep{ostorero2017}.

On the other hand, by increasing the column density, the total cloud size increases together with the line luminosity, which is demonstrated in Figs.~\ref{fig:rad_10pc_scales} and~\ref{fig:rad_50pc_scales}, where the single cloud structure is presented for the high Cen~A luminosity case and low starting density $n_{\rm H_0}=3.7 \times 10^{-4}$~cm$^{-3}$ for two distances from the SMBH:\@ 10 and 50 pc respectively. Different line styles represent cases with increasing column density, and clearly demonstrate the process of adding more material on the back side of the cloud. This material is becoming cold, dust-dominated and radially more extended. In each case, the outer size of the cloud is marked by a vertical line, and integrated CO line luminosity is shown.

Both figures show how the molecular gas layers are formed on the back, non-illuminated side of the clouds. In the case when the illuminated cloud surface is located at 10~pc, the molecular layer starts at 269~pc and it has a geometrical extension equal to 155~pc. But the estimated CO line luminosity is one order of magnitude below the observed value. The overall properties of such a cloud and computed line luminosities are presented in Tab.~\ref{tab:almalines} in the sixth column.

The situation changes when the cloud is located at 50~pc from the SMBH.\@ Fig.~\ref{fig:rad_50pc_scales} shows the extension of dust for this case, when the observed $L_{\rm CO}$ luminosity is almost fully reproduced. The modelled luminosities of several lines for this model are listed in Tab.~\ref{tab:almalines} in the seventh column. We found that in order to reproduce the observed CO line emissivity, the thickness of dusty material should be about 420~pc ($22\arcsec$), located around 1000~pc ($55\arcsec$). In both cases of clouds considered here, the hot plasma extends within 180~pc, which agrees with the {\it Chandra\/} observations.

The above results prompted us to propose a 3-D model of the hot plasma and molecular gas distribution, which is consistent with the {\it Chandra\/} observations and with the CO line observed by ALMA\@. Assuming a spherically symmetric model, the hot plasma, with temperature above $10^6$~K, is located in the innermost region ($10\arcsec$ diameter), which agrees with our data presented in Fig.~\ref{fig:contour}. With increasing radial distance the temperature of matter decreases and dusty molecular material starts to be visible. When projected onto the sky, we obtain a line luminosity similar to that observed within Reg 1, but the most dust concentration seen in ALMA data is located farther out in Reg 3 ($15\arcsec$ from the SMBH) and Reg 4 ($32\arcsec$ from the SMBH).

\begin{figure}
  \includegraphics[width=8.5cm]{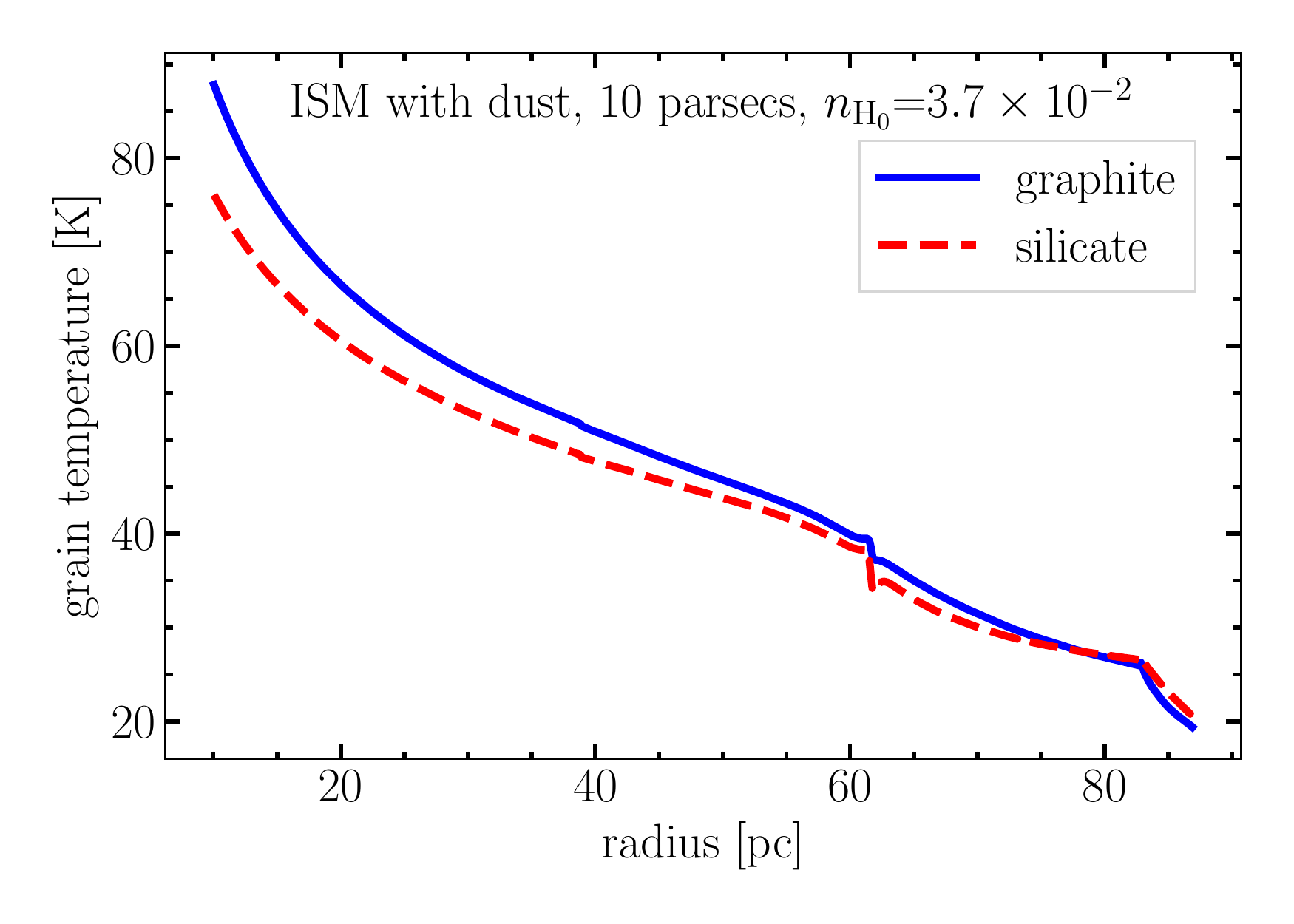}
  \caption{The temperature structure for two dust grains, graphite and silicate, for the model with initial density  $n_{\rm H_{0}}$= $3.7 \times 10^{-2}$~cm$^{-3}$ and for an inner radius of 10~pc.}
\label{fig:grains}
\end{figure}

\section{Discussion and Conclusions}
\label{sec:dis}

We analysed {\it Chandra\/} X-ray observations and ALMA data covering the mm band, focusing on the inner 10$\arcsec$ region (180 pc) of the Cen~A AGN/NGC~5128 galaxy system. The images show an overlap of the hot X-ray emitting plasma with the cold CO-emitting and dusty material. Our goal was to model the interaction of the hot plasma and the cold material and to verify whether TI leads to spontaneous formation of this multi-phase environment.

In order to understand the relevant processes we used one SED of broad-band emission from the very inner region of Cen~A available in the literature. We considered two luminosity states that differ by a factor of 70, which accounts for the fact that some emission may originate from the non-thermal jet and may be strongly boosted, and for the effect of intrinsic variability of the source. We performed our photoionization calculations under the assumption of all clouds satisfying the constant pressure approximation and the gas including heavy elements with ISM and Solar abundances.

We demonstrated that thermal instability is indeed able to generate the multi-phase medium, where the hot X-ray phase with temperature $\sim > 10^6 $ K stays in pressure equilibrium with the partially ionized gas of temperature $\sim 10^4 $ K that emits the H$_{\alpha}$ spectral line. Nevertheless, the cold dust of temperature $T < 100 $~K, responsible for the CO emission detected by ALMA, cannot co-exist with the X-ray emitting plasma but it could co-exist with the H$_{\alpha}$ emitting gas at $\log \Xi = -5$. This is a factor $10^5$ difference in the allowed level of irradiation. However, we note that this statement is correct for optically-thin gas, which is also the assumption behind the computations of our equilibrium curves.

With the advanced photoionization {\sc cloudy} computations we were able to search systematically over a wide parameter space. In this way we found the conditions for clouds of relatively low density, which can reproduce the observed CO line emission that reaches $1.51 \times10^{38}$ erg s$^{-1}$. This flux is achieved for clouds located at $\sim50$~pc and farther away from the nucleus. The modeled luminosities of all other observed lines (see Tab.~\ref{tab:almalines}) are higher than the observed ones, which we explain by the effect of shielding. Next, for the cloud located closer to the nucleus, the modeled CO line is one order of magnitude lower than the observed one, but the other lines are again overestimated.

The CO line is the most important from the observational point of view, and we thus concentrate towards reproducing its modelled luminosity. Note that such a line may only be produced in the high-luminosity state of Cen~A, since CO line luminosity decreases as the bolometric luminosity decreases.

From our modelling, we have demonstrated that the best distribution of matter that correctly reproduces the data is a 3-D cloud where the hot plasma extends up to $r_{\rm max} \simeq 200$~pc from the centre, which agrees with {\it Chandra\/} observations within $10''$. When the temperature drops below $1 \times 10^6$~K, we cannot consider this environment as a hot X-ray emitting medium. Instead, we start seeing the emergence of H$\alpha$-emitting clouds due to TI at $r_{\rm min} \simeq 200$~pc in the dust-less region, and $r_{\rm min} \simeq 1020$~pc in the dusty region (for the cloud of starting density $n_{\rm H_0}=3.7 \times 10^{-4}$~cm$^{-3}$ located at 50 pc). Nevertheless, this innermost part of Cen~A cannot be observed in the optical band since the dust lanes block the view. HST images reveal only outer filament structures, where the jet, whose diffuse X-ray emission is seen in the top panel of fig.~\ref{fig:contour}, induces star formation by kinematic interaction \citep{crockett2012}, thus we cannot test observationally the predictions for the ratios of the optical lines to CO and other molecules in the innermost region of our consideration.

A fraction of $1 \times 10^4$~K material is associated with the jet, and this means that we do not expect dust to be present in the jet region. Dust was probably destroyed in the outflow in the course of the accretion process. Therefore, the dusty material is a part of the spherical shell located outside the jet viewing angle; it should have a geometrical thickness of $\sim$420~pc, and it should be placed at $\sim$1000~pc from the nucleus. Such a location implies that we see the very inner region of Cen~A as a projected image on the sky through the dusty material. The location of the dust lanes further outside the nucleus is confirmed by the ALMA total map presented in Fig.~\ref{fig:regions}, where the most luminous CO line emission is found up to $50\arcsec$ on the northwest side, and up to even $60\arcsec$ on the southeast side of the disk-like shape. On the other hand, the mid-IR   high-resolution data from VLTI show the presence of an unresolved synchrotron component, and an extended dust disk with a projected size of order of 0.6 pc, with temperature of order of 240~K \citep{meisenheimer2007}. This observation does not disagree with our model, but still our theoretical modelling does not take into account the process of dust evaporation in the XDR, therefore we cannot say how long with time the dust grains can survive. Further consideration of dynamical modelling is needed to fully solve the existence of dust in XDR around SMBH.\@

Our choice of {\sc cloudy} software for modelling did not allow for deep and focused analysis of the molecular lines as could be done by using specialized software, for example radiative transfer codes like RADEX \citep{tak2007,kawamuro2019,kawamuro2020}, or the KOSMA-tau code \citep{stutzki1998}, which combines the modelling of individual spherical clumps in an isotropic UV field assuming a fractal structure for the ISM.\@ Some upgraded versions allow the user to even include cosmic ray heating and mechanical heating by turbulence and shocks or X-ray interaction with a partially ionized dusty medium \citep[e.g.][]{ms2005,ms2006,ms2007,kazandjian2012,kazandjian2016}. On the other hand, these codes do not consider X-ray-emitting hot plasma where Comptonization plays a dominant role, thus only {\sc cloudy} allows us to cover the complete temperature range from tens of K to $10^8$ K, i.e.\ to the temperature saturation at the inverse Compton temperature value for a given spectral shape of the incident continuum.

Our results confirm the existence of a multi-phase medium in Cen~A, including the hot X-ray emitting phase, as shown previously in two other sources: Sgr~A* and M60-UCD1, the latter being an ultra-compact dwarf galaxy \citep{czerny2013c, rozanska14, kunneriath2014, rozanska17}. We postulate here that active galaxies have enough power to produce a strong radiation field which then photoionizes the gas around their SMBHs. Such illumination enhances the classical TI, where under the usual constant pressure assumption the hot and cold gas can coexist. Dusty gas is moved outside the nucleus, most probably due to strong evaporation at the hot phase. The location of these phases agrees with currently available high-resolution images by X-ray and infrared telescopes. As the TI develops, it helps cold clumps to survive within the surrounding hot medium and drives them toward the central SMBH \citep{barai11, barai12}, thus enhancing the mass accretion rate during episodes of clump disruption and inflow.

\section*{Acknowledgments}
This paper is based on the following ALMA data: ADS/JAO.ALMA\#2015.1.00483.S. ALMA is a partnership of ESO (representing its member states), NSF (USA) and NINS (Japan), together with NRC (Canada) and NSC and ASIAA (Taiwan) and KASI (Republic of Korea), in cooperation with the Republic of Chile. The Joint ALMA Observatory is operated by ESO, AUI/NRAO, and NAOJ.\@ The research was supported by Czech Science Foundation No.~19-01137J (VK) and No.~19-05599Y (AB and PB), and the EU-ARC.CZ Large Research Infrastructure grant project LM2018106 of the Ministry of Education, Youth and Sports of the Czech Republic (AB). The research was also supported by Polish National Science Center grants No. 2015/17/B/ST9/03422, 2015/18/M/ST9/00541 (TPA and AR), 2016/23/B/ST9/03123, 2018/31/G/ST9/03224 (AGM), 2017/26/A/ST9/00756 (BCZ). BDM acknowledges support from the European Union's Horizon 2020 research and innovation programme under the Marie Sk\l{}odowska-Curie grant agreement No. 798726. TPA acknowledges the Nicolaus Copernicus Astronomical Center of the Polish Academy of Sciences (NCAC PAS) for its hospitality during his stay in Warsaw. We thank the anonymous referee for their constructive comments, which improved this paper.

\section*{Data Availability}
The data underlying this paper are available at http://almascience.nrao.edu/aq under ID: 2015.1.00483.S and https://cda.harvard.edu/chaser/ under ID: 20794. References to historical lightcurve data are provided in sec.~\ref{data:nustar}.

\bibliographystyle{mnras}
\bibliography{refs}
\label{lastpage}
\end{document}